\documentclass[twocolumn,trackchanges]{aastex7}
\usepackage{longtable}
\usepackage{threeparttable}

\newcommand\BV{Brunt-V\"ais\"al\"a}
\newcommand\ombv{\omega_{\rm BV}}
\newcommand\deltar{\delta r}
\newcommand\vkepler{v_{\rm K}}
\newcommand\drmax{\delta r_{\rm max}}
\newcommand\vrmax{v_{r, \rm max}}
\newcommand\sigv{\sigma_{v}}
\newcommand\tdiss{t_{\rm turb}}
\newcommand\leddy{l_{\rm b}}
\newcommand\nelec{n_{\rm e}}
\newcommand\edot{\epsilon}
\newcommand\keddy{k_{\rm b}}

\begin{document}

\title{Are X-ray Atmospheres Heated by Turbulent Dissipation? XRISM Constraints}

\author{B.R. McNamara}
\affiliation{Department of Physics and Astronomy, University of Waterloo, Waterloo Centre for Astrophysics, Waterloo, ON N2L 3G1, Canada}
\email{mcnamara@uwaterloo.ca}

\author[0000-0002-9378-4072]{A.C. Fabian}
\affiliation{Institute of Astronomy, Cambridge University, Madingley Rd., Cambridge, CB3 0HA, UK}
\email{acf@ast.cam.ac.uk}

\author[0000-0001-5208-649X]{H.R. Russell}
\affiliation{School of Physics \& Astronomy, University of Nottingham, Nottingham, NG7 2RD, UK}
\email{Helen.Russell@nottingham.ac.uk}

\author[0000-0003-0297-4493]{P.E.J. Nulsen}
\affiliation{ICRAR, University of Western Australia, 35 Stirling Hwy, Crawley, WA 6009, Australia}
\affiliation{Center for Astrophysics $|$ Harvard \& Smithsonian, 60 Garden Street, Cambridge, MA 02138, USA}
\email{paulnulsen@gmail.com}

\author[0000-0002-9714-3862]{A. Simionescu}
\affiliation{SRON Space Research Organisation Netherlands, Niels Bohrweg 4, Leiden, South Holland 2333 CA, The Netherlands}
\affiliation{Leiden Observatory, Leiden University, Niels Bohrweg 2, Leiden, South Holland 2333 CA, The Netherlands}
\affiliation{Kavli Institute for the Physics and Mathematics of the Universe, The University of Tokyo, Kashiwa, Chiba 277-8583, Japan}
\email{A.Simionescu@sron.nl}

\author[0000-0002-3525-7186]{A. Majumder}
\affiliation{Department of Physics and Astronomy, University of Waterloo, Waterloo Centre for Astrophysics, Waterloo, ON N2L 3G1, Canada}
\affiliation{SRON Space Research Organisation Netherlands, Niels Bohrweg 4, Leiden, South Holland 2333 CA, The Netherlands}
\email{anwesh.majumder@uwaterloo.ca}

\author[0000-0002-3031-2326]{E. D.\ Miller}
\affiliation{Kavli Institute for Astrophysics and Space Research,
Massachusetts Institute of Technology, 70 Vassar St, Cambridge, MA 02139}
\email{milleric@space.mit.edu}

\author[0000-0002-5222-1337]{A. Sarkar}
\affiliation{Department of Physics, University of Arkansas, 825 W Dickson st.,
Fayetteville, AR 72701, USA}
\affiliation{Kavli Institute for Astrophysics and Space Research,
Massachusetts Institute of Technology, 70 Vassar St, Cambridge, MA 02139}
\email{arnabs@uark.edu}


\begin{abstract}
 
We evaluate whether dissipation of turbulence injected into hot cluster atmospheres by jets and bubbles can offset radiative cooling flows. No trends are found between atmospheric velocity dispersion, $\sigma_v$, and either the ratio of kinetic to thermal energy or jet power over nearly four decades of jet power. Apparently, jets disperse their energy gently at  roughly constant energy per gram of gas. Assuming the velocity dispersions at the centers of Perseus, Virgo, and Hydra A reflect jetted turbulence, up to roughly half the bubble enthalpy could be dissipated by turbulent motion. 
A model is presented that balances radiation losses and turbulent power injected by radio bubbles rising at their terminal speeds. The model is anchored by XRISM measurements of $\sigma_v$ and is governed by the ratio of the bubble's terminal speed to the atmospheric sound speed. Bubbles must rise close to the sound speed and impart energy with a broad range of injection scales to heat the entire cooling volume. The level of turbulence in the powerful Hydra A system may offset cooling over some of the cooling volume.  However, turbulent dissipation would struggle and probably fail to balance cooling in Perseus and Virgo, except perhaps in their inner regions. Several factors including, low velocity dispersions, small injection scales, short duty cycles, anisotropic turbulence injection, and long turbulent diffusion timescales present severe challenges for jetted turbulence heating models.  A larger sample of spatially resolved cluster atmospheres is needed to reach a definitive conclusion.
\end{abstract}



\section{Introduction} \label{sec:Intro}

X-ray images from the Chandra Observatory have shown that dense, cooling atmospheres pervading galaxies and galaxy clusters usually contain cavities several to tens of kpc across straddling the central galaxy.  The cavities are usually filled with radio synchrotron light indicating that they were inflated by powerful radio jets.  After they are inflated, the cavities rise through the atmospheres like bubbles propelled by buoyancy \citep[][]{2000Churazov,McNamara2000,Churazov2001}. The jet power derived from their enthalpy and ages is comparable on average to the X-ray luminosity emitted from the bright central regions of cluster atmospheres \citep[][]{Birzan2004}. Jets therefore have enough power to prevent most of this gas from cooling to lower temperatures and forming a cooling flow \citep[][]{Rafferty2006, Dunn2006}.  While the atmospheres are thermally stable overall, pockets of thermally unstable gas near the central galaxy cool and condense into reservoirs of molecular clouds \citep[][]{Olivares2019,2019Russell} that fuel low levels of star formation \citep[][]{McDonald2018}.  A small fraction of the molecular clouds feed the central galaxy's nuclear black hole \citep[][]{2024Oosterloo} and sustain a feedback loop  \citep[reviewed by][]{McNamara2007, McNamara2012,2022DonahueVoit}.  These and other studies \citep[e.g.,][]{2003Fabian,2005Forman,2005Voit} were foundational to the current view that radio jets are the primary agent that quenched star formation in massive galaxies following Cosmic Noon \citep[][]{2023Heckman} and have since prevented significant levels of star formation in contemporary massive galaxies \citep[][]{2014Heckman,2006Bower,2006Croton}. 


While jets are powerful enough to thermally stabilize hot atmospheres \citep{2025Igo}, how their energy is dissipated and how they achieve thermal balance over enormous scales is a more challenging problem.  Mechanisms proposed to achieve this balance include mixing between the relativistic particles within the bubbles and the surrounding thermal plasma \citep[][]{Hillel2016}, cosmic ray heating \citep[][]{2023Ruszkowski,2008Guo,2008Colafrancesco,Fujita2013}, shock and sound wave heating \citep[][]{Nulsen05,Randall2015,2018Zweibel, 2003Fabian,Fabian2017}, circulation flows \citep[][]{2003Matthews,2021Bourne}, dissipation of jetted turbulence \citep[][]{Zhuraleva2014, 2005Dennis}, or a combination of these mechanisms.

 Heating by the dissipation of atmospheric turbulence is an appealing candidate.  X-ray cavities are ubiquitous in bright cluster cores. The lifting and displacement of hot gas by the bubbles presumably produces turbulence as the bubbles inflate and rise through the atmosphere.  
 
 Emission line width measurements of 62 cluster atmospheres using XMM-Newton's Reflection Grating Spectrometer yielded upper limits of $\sigma_v < 500~\rm km~s^{-1}$ for most systems and less than $200~\rm km~s^{-1}$ for a few bright nearby systems \citep[][]{Sanders13}. These measurements, with precision limited primarily by the finite sizes of cluster cores, lie well above the velocity dispersion measured for the Perseus cluster atmosphere by the Hitomi observatory \citep[][]{Hitomi18}.

X-ray surface brightness fluctuation measurements have yielded more interesting limits on the turbulent velocity field in cluster atmospheres \citep[][]{2012Churazov, 2012Sanders,Zhuraleva2014, Li2025}. The interpretation of these difficult measurements assumes the fluctuations are produced primarily by adiabatic gas motions propelled by buoyancy. If the fluctuations are associated with gas density variations induced by turbulent motion, they would be sensitive to the velocity power spectrum on large scales from which a turbulent dissipation rate can be inferred. 

Spectral slopes were found to be broadly consistent with a Kolmogorov spectrum that in turn yielded dissipation rates broadly consistent with cooling X-ray luminosities \citep[][]{Zhuraleva2014}.  However, large measurement uncertainties allow for a broad range of slopes \citep[][]{2012Sanders,Li2025}.  
Moreover, surface brightness fluctuation amplitudes in low jet power systems are similar in amplitude to strongly jetted atmospheres \citep[][]{2012Sanders,Li2025} indicating significant power from large-scale turbulence that would be unable to dissipate on timescales short enough to quench cooling. 


Velocity fields of the cooling and cold gas have been probed using nebular emission from atomic hydrogen and CO emission from molecular hydrogen. Nebular and CO emission have the great advantage that they probe velocities along many sight lines with high angular resolution on scales of a few kpc or less.  These scales generally lie below jet and bubble sizes allowing the velocity spectrum to be probed in detail.   

Nebular and CO velocity structure functions yield turbulent speeds between a few tens to a few hundred kilometers per second \citep[][]{2018Gaspari, Li20, Mohapatra2019,Mohapatra2022, Li2025}.  On the largest measurable scales of a few tens of kpc, nebular velocities approach the turbulent speeds found by Hitomi and XRISM \citep[][]{Li20,Li2025}.  This correspondence may indicate that the cool and hot gas phases in some instances may share similar kinematics on large scales despite vast differences in density and filling factors. 
Slope changes or turnovers in the nebular and CO velocity power spectra often occur on spatial scales comparable to the bubble sizes, suggesting the gas turbulence is injected by the rising bubbles \citep[][]{Li20,Li2025}.
 Lower velocity dispersions found beyond the bubble regions may perhaps reflect calmer atmospheres away from the bubbles \citep[][]{Gingras2024,2024Vigneron}.

Nebular and molecular gas velocity spectra yield steeper spectral slopes than the Kolmogorov scaling expected for gaussian random turbulence.  This steepening may be caused by gravitational acceleration of the dense clouds \citep[][]{Li2025,2021Wang} and/or viscosity and magnetic stresses between the clouds and the surrounding atmosphere \citep[][]{Mohapatra2022,Fournier2024}. If the nebular and molecular VSFs mirror those of cooling X-ray atmospheres, the steep spectra would indicate inefficient turbulent heating \citep[][]{2023Ganguly}.  However, the mean densities and volume filling factors between the colder clouds and hot atmospheres differ vastly, and the hot and cold gas phases are decoupled on some scales \citep[][]{Li2025}. 

Fresh insight into atmospheric turbulence is emerging from observations of cluster atmospheres with the XRISM observatory's Resolve microcalorimeter detector \citep[][]{2025Tashiro}. Resolve has achieved roughly 5 eV spectral resolution near the 6.7 keV iron line complex, albeit with low $\approx 1.5$ arcmin angular resolution.  The instrument is yielding atmospheric peculiar velocities and velocity dispersions with precisions of tens of km per second. Early results indicate ratios of non-thermal to thermal pressures of only a few percent or less \citep[][]{XRISM_2025_A2029,2025Rose,XRISM_Ophiuchus2025}, which appear to be in tension with cosmological models \citep[][]{XRISMCosmology,2026Vazza}.

We investigate here whether turbulence indicated by the velocity widths of atmospheric spectral lines, the 6.7 keV iron complex foremost among them, is sufficiently large to offset radiative cooling. We apply simple, ideal models to explore the conditions required for turbulent heating to offset cooling in several representative systems, and we discuss whether or not they are physically plausible. We assume a standard $\Lambda$CDM cosmology with $H_0 = 70$ km s$^{-1}$ Mpc$^{-1}$, $\Omega_{\Lambda} = 0.7$ and $\Omega_m = 0.3$.

\section{Data} \label{sec:Data}

The atmospheric velocity dispersions for an archival sample of ten clusters were taken either from published XRISM PV analyses or from GO data as indicated here and in the text: M87 \citep[][]{2025XRISM_M87}, Perseus \citep[][]{2025XRISM_Perseus}, PKS0745-191 \citep[][]{2026Tanaka}, Abell 2029 \citep[][]{XRISM_2025_A2029}, Centaurus \citep[][]{2025XRISM_Centaurus}, Cygnus A \citep[][]{2025Majumder}, Hydra A \citep[][]{2025Rose} and Majumder et al. (2026 in preparation), Abell 1795 (Sarkar et al. 2026, ApJ, submitted), Coma \citep[][]{XRISM2025Coma}, Ophiuchus \citep[][]{XRISM_Ophiuchus2025}. The mass and luminosity profiles were taken from the data base developed by \citet[][]{Pulido2018} and \citet[][]{Hogan2017}. Unless otherwise noted, the
jet powers are primarily from \citet[][]{Rafferty2006,Dunn2006, Birzan2004,Wise2007} throughout. For Abell 2029, which has no discernible X-ray cavities, the synchrotron power was scaled by the average jet power scaling relations of \citet[][]{Birzan2008} and \citet[][]{2010Cavagnolo}.

\section{Analysis}\label{sec:Analysis}

\subsection{Do Jets \& Bubbles Generate Atmospheric Turbulence?}

XRISM has already probed jet power from several $\times 10^{42}~\rm erg~s^{-1}$ in the Centaurus cluster \citep[][]{2025XRISM_Centaurus} to greater than $10^{46}~\rm erg~s^{-1}$ in Cygnus A \citep[][]{2025Majumder}.  If jets and bubbles generate turbulence at levels that scale with jet power, we would expect a trend between jet power and central atmospheric velocity dispersion.  Figure 1 shows no trend over nearly four decades of jet power.  While the highest velocity dispersion is seen in Cygnus A, the central dispersion in M87 is comparable despite a jet power two decades below.  In all likelihood, the high central velocity dispersions in both systems reflect unresolved bulk motions along the line of sight rather than fully-developed turbulence \citep[][]{2025Majumder,2025XRISM_M87}. 

Spatial resolution is at issue in Figure 1.  M87's velocity dispersion profile does not decline smoothly despite high spatial resolution.  It's nuclear $\sigma_v\sim 260~\rm km~s^{-1}$ within a 5 kpc radius plummets to $60~\rm km~s^{-1}$ beyond.  A similarly strong jump in more distant systems would be diluted by XRISMs limited angular resolution.  Cygnus A's similarly large velocity dispersion is measured over 200 kpc, reflecting vastly larger turbulent energy than M87 despite similar velocity dispersions. Apparently, jets and bubbles disperse their energy gently at a roughly constant energy per gram of gas, which we explore further below. 
\begin{figure}
    \centering
    \includegraphics[width=\columnwidth]{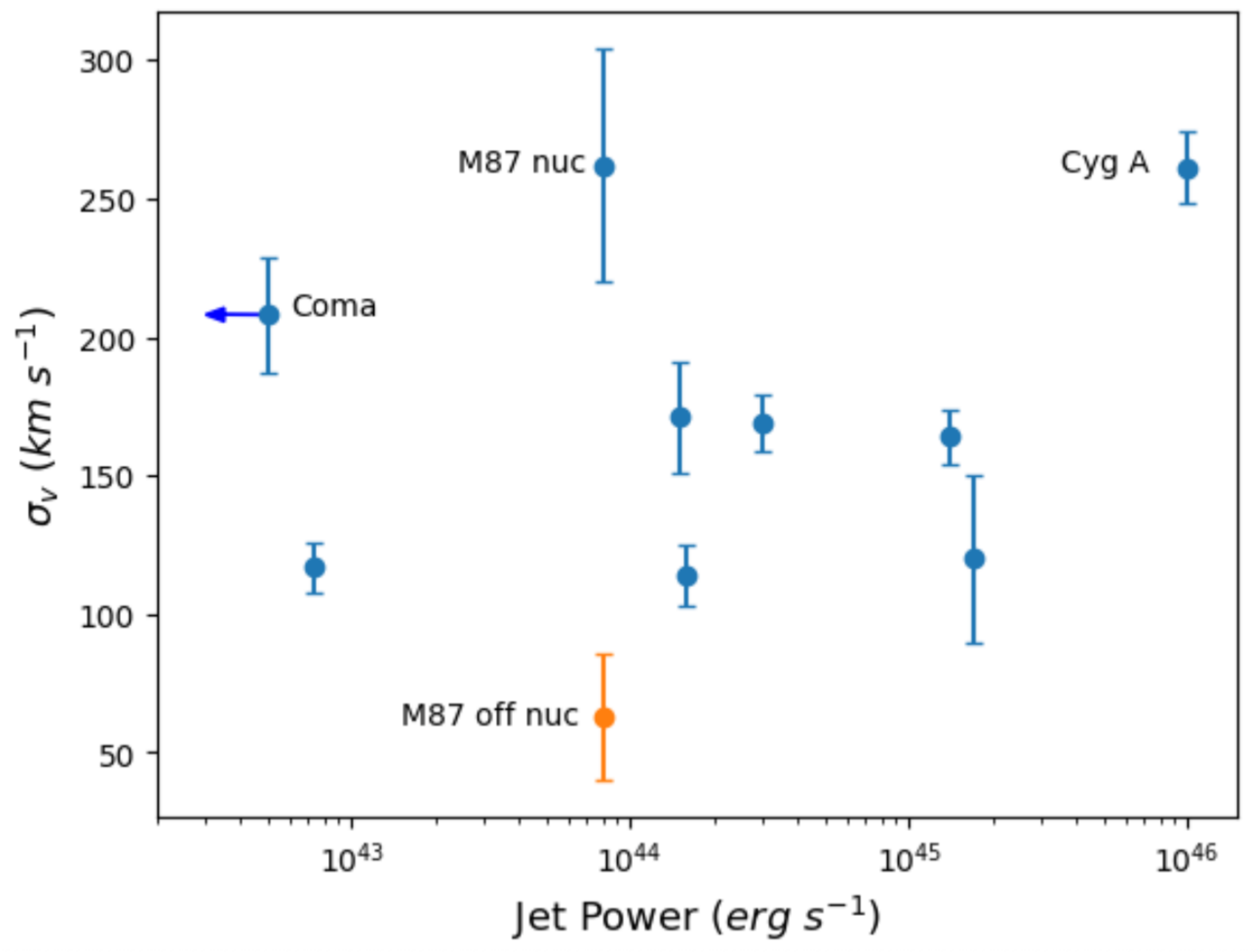}
      \caption{Central velocity dispersion is plotted against Jet power based on data from the literature.  No trend is found.}
\end{figure}

\begin{figure}
    \centering   
    \includegraphics[width=\columnwidth]{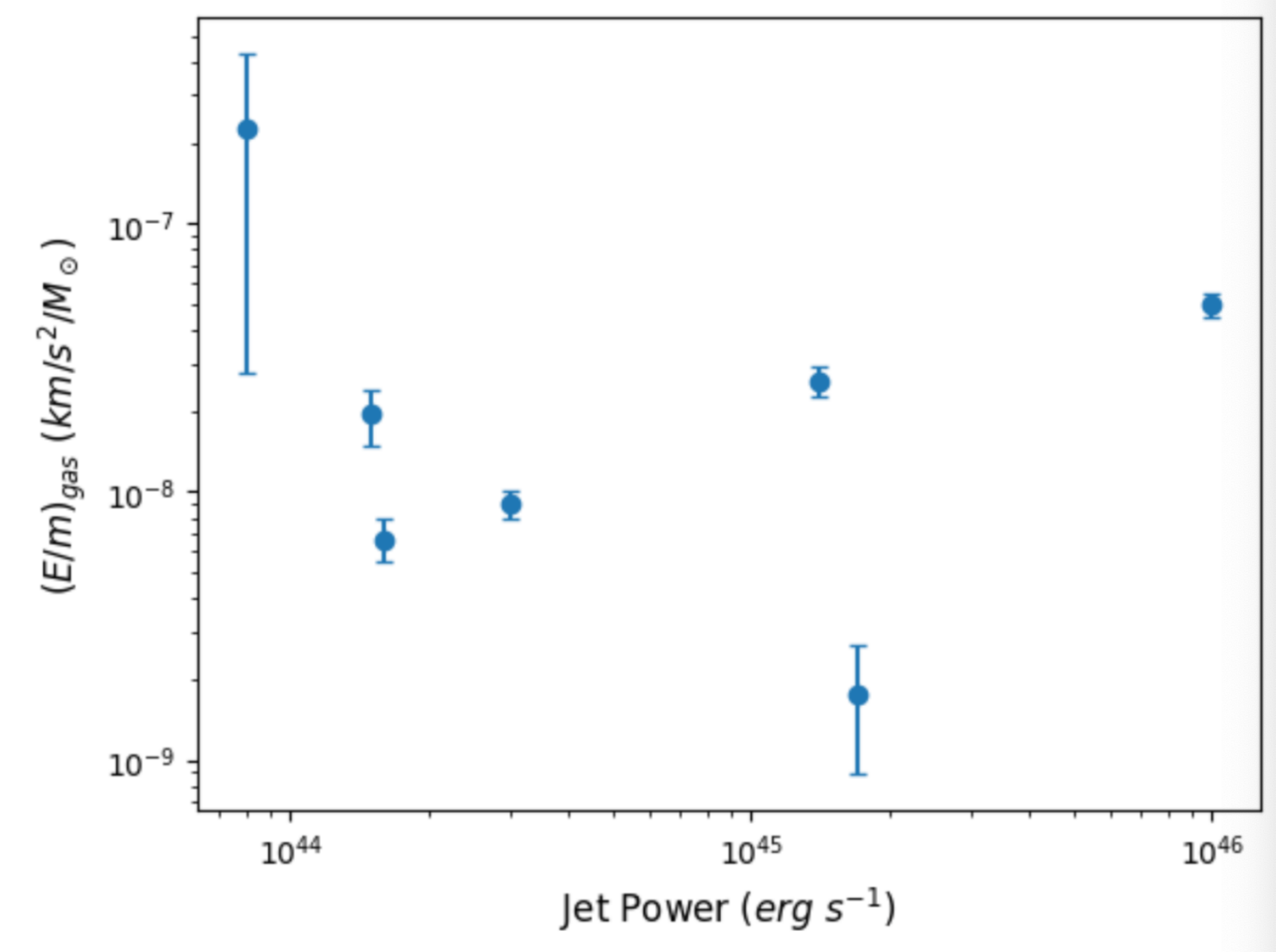}
    \caption{Gas specific energy, or atmospheric energy per unit mass plotted agains jet power for seven systems with measured gas mass profiles.  No trend is apparent. }
\end{figure}
\begin{figure}
    \centering   
    \includegraphics[width=\columnwidth]{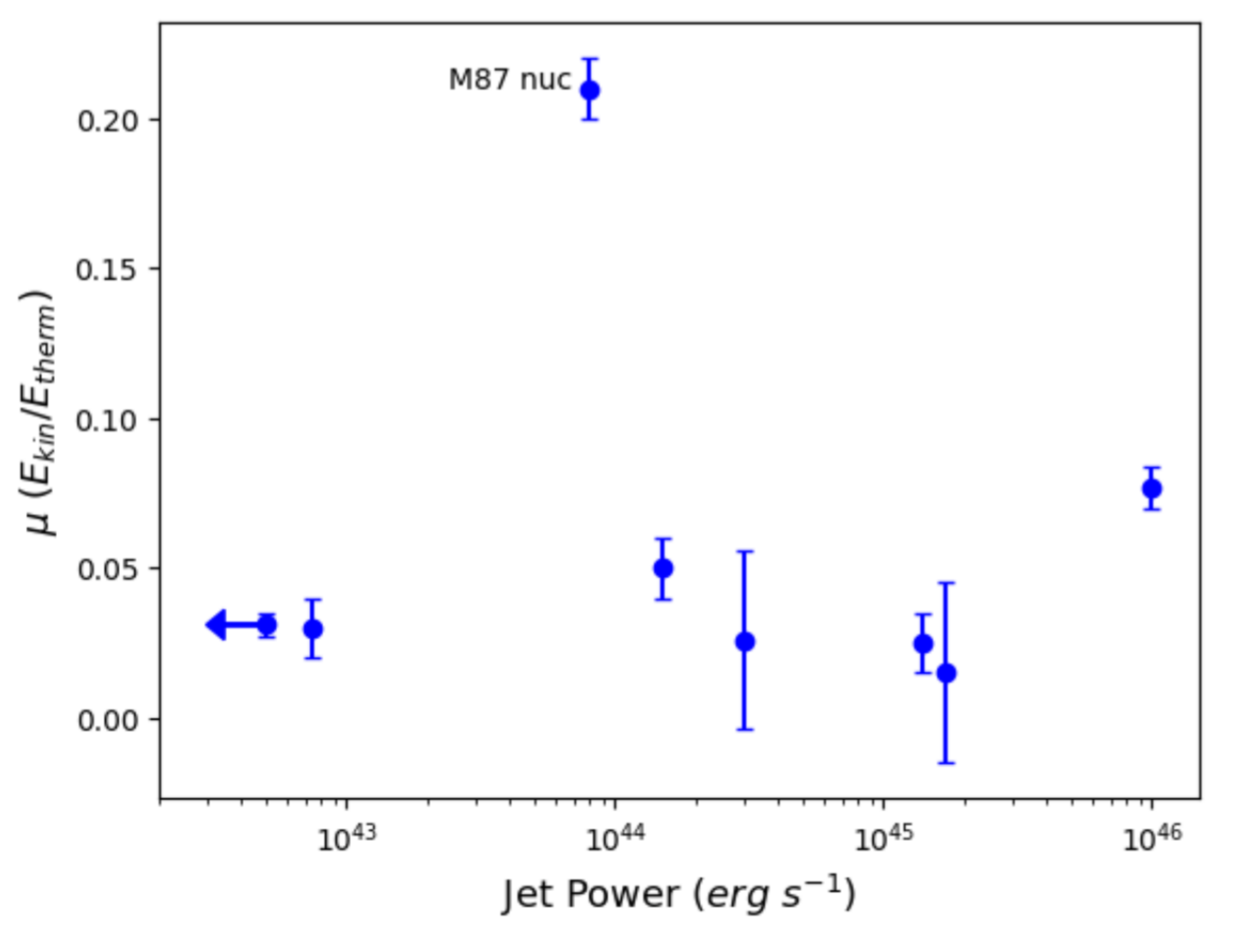}
    \caption{Jet power plotted against the ratio of kinetic to thermal energy.  No trend is apparent. }
\end{figure}
Jet power tends to correlate with the size of the jets and lobes. Higher power jets dissipate their energy over larger volumes and larger atmospheric gas masses \citep[][]{2008Diehl}.  A jet of a given power dissipating its energy into a larger gas mass would have a lower turbulent speed.  Therefore, specific turbulent energy (energy per unit mass) within the XRISM pointing is plotted against jet power for seven systems with gas mass measurements available to us in Figure 2.  This approach effectively integrates approximately over the effective length scale discussed in detail by \citep[][]{2025Rose,2025XRISM_Perseus,2025XRISM_M87}. No correlation is evident.  

\begin{table}[ht]\centering
\begin{threeparttable}
\caption{Central Velocity Dispersions}
\label{tab:my_table}
\begin{tabular}{lll}
\hline
Object & $\sigma_v$ ($\rm km~s^{-1}$)& Ref \\
\hline
M87     & $262\pm 45$& [1] \\
Perseus & $171\pm 20$ & [2]\\
PKS 0745-191& $120 \pm 30$ & [3]\\
Abell 2029& $169\pm 10$ & [4] \\
Centaurus &  $117\pm 9$& [5]\\
Cygnus A& $261\pm 13$ & [6]\\
Hydra A& $164\pm 10$ & [7]\\
Abell 1795 & $110 \pm 10$ & [8] \\
Ophiuchus & $115 \pm 12$ & [10] \\
\hline
Mean & $165 \pm 60$ &...\\
\hline
Coma & $208 \pm 21$ & [9] \\
\hline

\end{tabular}
\begin{tablenotes}
\small

    \item [1] \citet{2025XRISM_M87}
     \item [2] \citet{2025XRISM_Perseus}
      \item [3] \citet{2026Tanaka}
      \item [4] \citet{XRISM_A029} 
      \item [5] \citet{2025XRISM_Centaurus}
      \item [6] \citet{2025Majumder} 
      \item [7] \citet{2025Rose} 
      \item [8] Sarkar et al. (2026, Submitted)
      \item [9] \citet{XRISM2025Coma}
      \item [10] \citet{XRISM_Ophiuchus2025}

\end{tablenotes}
\end{threeparttable}
\end{table}

Similarly, in Figure 3 we plot the ratio of kinetic to thermal energy, ${E_k / E_{th}} = \mu m_p \sigma^2 /kT$, against jet power. 
This ratio 
should be sensitive to energy added by jets to preexisting thermal energy.  No trend is apparent. 
Figures $1-3$ indicate to the degree that jets inject turbulence, they do so by gently and widely distributing the energy.

\subsection{Radial Dependence of \texorpdfstring{$\sigma_v$}{sigma\_v}}

Understanding the degree to which turbulent dissipation heats cluster atmospheres rests on disentangling gas motions generated by jets and bubbles from larger-scale motion generated by mergers.  Large-scale turbulent motions projected along the line of sight would have large injection scales and long dissipation time scales that would not heat effectively \citep[][]{Bourne2017,2025Li_Yang}.  Only a handful of systems are resolved well enough to address this question. 

The radial dependence of $\sigma_v$ within the cooling regions of seven systems is shown in Figure 4. The data points reflect the midpoints of the angular resolution bins, while the x-axis bars represent bin widths.  Broader bins indicate more distant systems with lower spatial resolution.  Systems with strong jetted motions should reveal rising dispersions near the jets and bubbles.  Figure 4 reveals inward-rising velocity dispersion only in M87 and Perseus.  

The highest dispersion is found in the inner 5 kpc of M87 which is probably bulk gas motion driven by the jets and shock front \citep[][]{2025XRISM_M87}.  Beyond the inner peak, M87's profile is flat between 5 kpc and 25 kpc.  
Perseus reveals the strongest evidence for jet- and bubble-related motion, with $\sigma_v$ peaking at the center and declining smoothly toward the cooling radius at 100 kpc \citep[][]{2025XRISM_Perseus}. Its velocity dispersion rises at higher altitudes (not shown) beyond the cooling region indicating significant large-scale motions \citep[][]{2025Bellomi}.  

Hydra A's $\sigma_v$ profile is flat within the $<120$ kpc cooling region and beyond to the northern cluster-scale cavity at 200 kpc (Majumder et al. 2026, in preparation). The XRISM footprint is filled by bubbles, so the flat $\sigma_v$ profile may indicate jetted turbulence throughout the footprint.  Coma contains no cooling flow and no radio source.  Its profile and large $\sigma_v$ are similar to Hydra A \citep[][]{XRISM2025Coma,2025Gatuzz}.   

The Ophiuchus cluster's velocity dispersion rises from $\sigma_v = 115 \rm ~km ~s^{-1}$ at the centre to $186~\rm ~km ~s^{-1}$ at roughly 40 kpc \citep[][]{XRISM_Ophiuchus2025}.  Ophiuchus apparently experienced an enormous jet outburst in the last 240 Myr \citep[][]{2020Giacintucci}, but has lain dormant since, indicated by a left arrow in Figure 3. Its radio-jets may have created turbulent motions in the past during its expansion and early buoyancy phases. Perhaps this motion has thermalized, was radiated away, or deposited at larger radii.  

The absence of a clear, inward-rising trend in Figure 4 may reflect, in part,  poor spatial resolution or longer effective length scales at larger radii that tend to elevate the outer dispersions. Simulations provide useful insight. For example, the rising velocity dispersion in Perseus is seen in a tailored $10^{45}\rm~erg~s^{-1}$ jet simulation with a similar velocity amplitude \citep[][]{2025Bellomi}. They found that larger-scale turbulence from sloshing contributes only 10\% to the central $\sigma_v$  \citep[][]{2025XRISM_Perseus}.  
Other simulations have found central gas motions elevated by projected, larger-scale motion along the line of sight \citep[][]{Bourne2017,2021Ehlert}, or transient, unresolved bulk flows rather than fully-developed turbulence \cite{2025Li_Yang}. The situation is unsettled.

\begin{figure}
    \centering
    
    \includegraphics[width=\columnwidth]{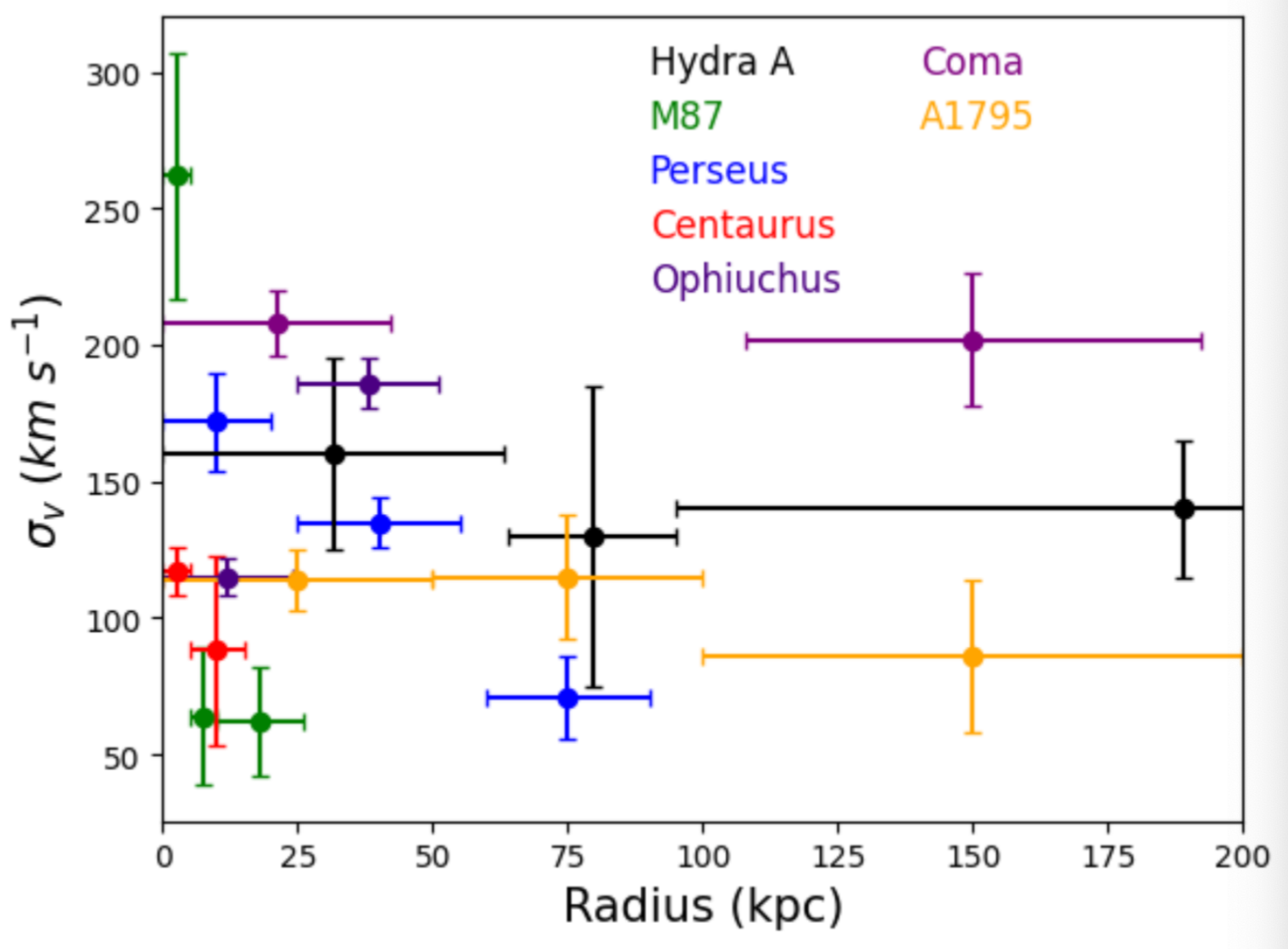}
    \caption{Atmospheric velocity dispersion plotted against radius or altitude for several systems with spatially-resolved velocity measurements within the cooling volume.  The points indicate the midpoint of the bin and the horizontal bars indicate the bin size.  The x-axis bar sizes indicate the relative spatial resolution of each system. }
    \label{}
\end{figure}

\subsection{Fraction of Bubble Enthalpy in Atmospheric Turbulence}

XRISM and Chandra observations together can place upper limits on the fraction of jet enthalpy dissipated by turbulence while the bubbles inflate. Hydra A's central footprint encloses a radius of about 95 kpc.  Cavities A through D, with total enthalpy $2.4\times 10^{60}~\rm erg$, lie within this radius \citep[][]{Wise2007}.  The enclosed atmospheric kinetic energy, assuming a one dimensional velocity dispersion of $160~\rm km~s^{-1}$, is $1.1\times 10^{60}~\rm erg$.  The upper limit on the conversion of jet enthalpy to kinetic energy is $\sim 46\%$, some fraction of which may be in turbulence.  Assuming the enthalpy of all six bubbles contribute, this figure drops to $\sim 20\%$.  A similar calculation for M87 reveals that roughly half the AGN outburst energy may be deposited in unresolved bulk motion and turbulence \citep[][]{2025XRISM_M87}. 

The atmospheric kinetic energy enclosing the inner bubbles of Perseus within 20 kpc is $\sim 7 \times 10^{58}~\rm erg$. The enthalpy released during their inflation phase is $\sim 1.4\times 10^{59}~\rm erg$ \citep[][]{Birzan2004}, indicating that roughly 50\% or less of the $4pV$ jet power may be converted to gas motions. However, if the total jet power lies closer to $10^{45}~\rm erg~s^{-1}$ \citep[][]{2003Fabian}, the fraction imparted in turbulence would lie below $20\%$.

These figures are approximations. They do not account for unresolved bulk motion superposed along the lines of sight that would inflate $\sigma_v$, nor do they fully account for power released through other channels such as shocks, sound waves, and cosmic rays which would lower the turbulent energy fraction.  Nevertheless, these figures indicate a significant fraction of jet power may be deposited in kinetic energy that may eventually thermalize and heat the inner regions of cluster atmospheres. 

Hydrodynamical models have shown that jet-inflated bubbles generate turbulence through the decay of azimuthal g-modes \citep[][]{2015Reynolds}.  However, fraction of jet enthalpy this process channeles into turbulence is only a few percent or less \citep[][]{2008Scannapieco_Bruggen,2015Reynolds,Bourne2017}.  
We find a higher fraction, although our measurements are upper limits.  

\subsection{Propagation of Turbulence}

Estimating the level of turbulent dissipation is complicated by many unknowns.  Jets and bubbles may inject energy primarily on a single scale or over a range of scales \citep[][]{HillelSoker2020}. The energy containing scale(s), $l$, must be comparable to the agent creating turbulence. Jets and bubbles span scales of a few to a few hundred kpc, and individual jets and bubbles grow with time. We therefore cannot measure injection scale(s) with precision. The turbulent dissipation timescale $\tau_{\rm turb}\sim l/\sigma_v$, which is roughly the turnover timescale for the largest eddy, is also poorly known. Cluster atmospheres are weakly magnetized which can affect the dissipation timescale.  

We measure the projected atmospheric energy within the effective volume \citep[][]{2025Rose},

\begin{equation}
E_{\rm atm} = 3\times 10^{59} \left({M_{\rm atm} \over 10^{12}\,\rm M_\odot}\right) \left({\sigma_v \over 100\,\rm{km\,s}^{-1}}\right)^2 ~\rm{erg},\
\end{equation}
assuming $\sigma_v$ represents pure isotropic turbulence.

Turbulent dissipation of jet power may be able to offset cooling provided, 1) the atmosphere's turbulent kinetic energy produced by jets and bubbles is sufficiently high, 2) the turbulent speeds and injection scales permit the energy to cascade to small scales and dissipate throughout the cooling volume before the energy is radiated away, 3) energy is replenished at a sufficient average rate with short enough duty cycle to offset cooling \citep[][]{Fabian2017, Bambic18,2025Rose}.  Jets are, on average,  powerful enough to do the job \citep[][]{Rafferty2006,2025Igo}, but the remaining requirements are at issue.  

Jetted turbulence must fill the cooling volume while being generated centrally, which is problematic.   Jets and bubbles generate turbulence locally and anisotropically, primarily around the lobes, terminals, and in the updraft \citep[][]{2021Ehlert,Bourne2017,2012Vazza_Roediger,2025Li_Yang}.  The jetted turbulence diffuses slowly away \citep[][]{2006Rebusco}.  Volume-filling turbulence is more readily produced by mergers \citep[][]{2012Vazza_Roediger,2025Bellomi}.  

\cite{Fabian2017} showed that turbulence generated centrally could not outpace radiation if turbulence diffused outward at the turbulent speed. Fabian's calculation was optimistic but makes the point. Turbulence would fill the volume at a much lower rate than the turbulent speed. \cite{2012Vazza_Roediger} found maximum large-scale diffusion rates of $D_{\rm turb}\sim 10^{29}-10^{30}~\rm cm^2~s^{-1}$ from mergers and sloshing, but much smaller values of $\sim10^{28}~\rm cm^2~s^{-1} $ from jets and lobes.  Li \& Yang (2025, private communication) found a diffusion coefficient $D_{\rm turb}\sim 7\times 10^{29}~\rm cm^2~s^{-1}$. At these rates it would take a few $10^9~\rm yr$ for turbulence to diffuse 10 kpc from its injection site.   Therefore, turbulence must be injected locally throughout the cooling volume. \cite{2025Rose} suggested bubbles rising at the terminal buoyancy speed may be a viable mechanism, which is investigated further below. 

\subsection{Relevant Timescales}

We adopt as an example 
Abell 2029 which has one of the highest X-ray cooling luminosities \citep[][]{Birzan2004,2026Watson} and at $2\times 10^8~\rm yr$ one of the shortest central cooling times known in a hot cluster \citep[][]{McNamara16}. Yet its central galaxy contains no detectable cooling gas in the forms of nebular line emission, molecular clouds, or star formation \citep[detailed in][]{2025Sarkar,2004Blanton}. The central galaxies in nearly all systems with thermodynamic properties similar to Abell 2029 burgeon with star formation at rates of tens to hundreds of solar masses per year and are awash with reservoirs of $\sim 10^{10}~\rm M_\odot$ of molecular gas and bright nebular emission.  
Instead, Abell 2029's central galaxy is an old giant elliptical.

Its atmosphere hosts a prominent sloshing spiral and other low-level structures associated with a gravitational fly-by \citep[][]{2004Blanton,2026Watson}. 
No X-ray cavities have been found despite hosting a relatively large and powerful radio source. Nevertheless, its cooling flow is nearly quenched. The bubbles normally associated with such a powerful radio source may have been destroyed by the sloshing motion \citep[][]{2022FabianZuhone}, which also led to its head-tail morphology. However, the buoyant radio lobes rising through the atmosphere may nevertheless generate turbulence. The stresses on any viable heating and cooling mechanism are extreme in Abell 2029, making it a key system to explore the demands on turbulent heating. 


Figure 5 shows in black the radial run of time required to radiate away thermalized turbulent energy, $\tau_{\rm rad}(r)=E_{\rm turb}(r)/L_{\rm x}(r)$, within the cooling region. The radiative cooling time, the time required to radiate away all of the atmosphere's thermal energy plotted as a dashed blue line, is nearly a factor of ten longer.  The sound crossing timescale required for centrally-generated sound waves and pressure disturbances to propagate outward is shown in red.  

The timescale for a bubble to traverse the atmosphere traveling at its buoyancy terminal speed, assumed here to be $0.5c_s$ \citep[][]{Churazov2001}, is shown in green.  We show below that turbulence injected by bubbles moving outward at half the sound speed would be able to balance radiative losses over some fraction of the cooling regions of some systems.  The timescale for turbulence to cascade and thermalize will vary depending on several factors including the injection scale $l$ and turbulent speed $\sigma_v$ as $t_{\rm turb} \sim l/\sigma_v$. The turbulent energy density is smaller than the thermal energy density, so $t_{\rm turb}$ must be much shorter than $t_{\rm cool}$.  Its specific value will depend on model assumptions including the rate of energy injection and whether or not this rate is balanced by radiation losses, which need not be strictly true.  Our models assume equilibrium between heating and radiation losses so that $t_{\rm rad}\simeq t_{\rm turb}$.



\begin{figure}
    \centering
    
    \includegraphics[width=\columnwidth]{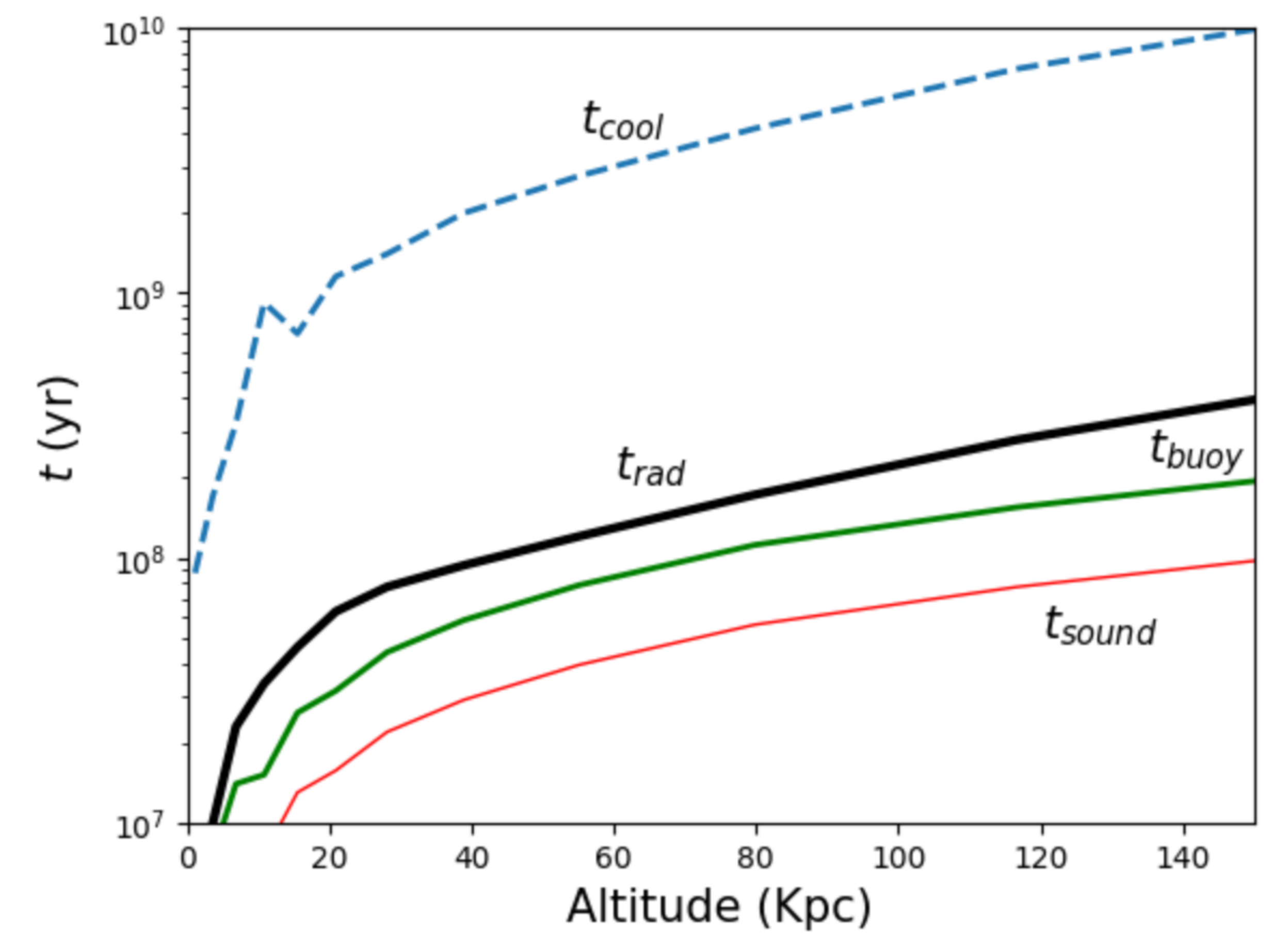}
    \caption{Relevant timescales as a function of radius using Abell 2029 as an example. The curves are explained in Section 3.5.}
\end{figure}


\subsection{Turbulent Dissipation of Jet Energy}
The heating rate due to turbulent dissipation with a Kolmogorov spectrum may be expressed as 
\begin{equation}
   \dot E\simeq  {E_{\rm atm}\over \tau_{\rm turb}} \propto {\sigma_v^3 \over l}, 
\end{equation}
 where $l$ is the injection scale, the energy-containing scale, or the outer scale. In this approximation, energy is injected throughout the cooling volume at the injection scale and cascades to smaller scales at a constant rate over the timescale $\tau_{\rm turb}\sim l/\sigma_{\rm turb}$ (we assume throughout that $\sigma_{\rm turb}=\sigma_v$). The Kolmogorov spectrum scales as $v \propto l^{1/3}$. Therefore, $\dot E$ is constant in the inertial range on spatial scales smaller than $l$.

A XRISM pointing alone cannot constrain $l$ and $\sigma_{\rm turb}(l)$ \citep[][]{ZuHone16,2025Rose}, nor can it measure the injection scale. The injection scale is normally assumed to be roughly the size of the agent injecting the energy, such as a radio jet, lobe, or bubble.  Adopting an injection scale without velocity measurements on or below that scale must be done deliberately.

This problem has been approached by assuming an effective length scale \citep[][]{2025XRISM_Perseus,2025Rose,2025XRISM_M87}, which is the largest measurable velocity scale per angular resolution element.  The effective length scale is usually estimated as the atmospheric depth along which an arbitrary fraction, say 50\%, of the light emerges. This scale is usually several tens of kpc which generally exceeds the sizes of bubbles and jets that inject the turbulence. The shortest effective scales are found in the bright cores where the jets and lobes reside.  The effective scale may be much larger than the injection scale leading to a large uncertainty in turbulent power estimates \citep[][]{2025Rose,2025XRISM_Perseus,2025XRISM_M87}.

We avoid this approach.  Instead, the measured velocity dispersions, $\sigma_v$, are used to constrain the injection scales and velocity dispersions required to successfully balance radiation losses.  The models are ideal and optimistic, as they assume the observed velocity dispersions reflect ideal, isotropic turbulence with a spectrum close to Kolmogorov scaling. They further assume $\sigma_v$ represents local values of the turbulent energy density in the cooling regions.  We do not account for foreground and background motions projected along the line of sight.  The energy densities measured here are upper limits. The injection and dissipation timescales are assumed to be in equilibrium, which need not be true in detail.  However, energy must be re-injected on timescales between $t_{\rm turb}$ and $t_{\rm cool}$ (Figure 5) to effectively offset cooling.  

With these caveats, statistical fits to the data would not be warranted.  Instead, the models are scaled to the data by matching the integrated luminosities to determine the ranges of $\sigma_v$ and $l$ that would be required to balance cooling under ideal conditions. We then ask whether they are physically plausible.

\subsection{Phenomenological Model of Turbulent Dissipation}

The turbulent power injected as a function of radius over the XRISM image is approximately:
\begin{equation}
  P_{\rm turb} \simeq {3 \over 2} M(r){\sigma_v^3 \over l}. 
\end{equation}
$M(r)$ is the atmospheric gas mass subtended by the image. The turbulent dissipation timescale, which is not well understood in a weakly-magnetized stratified atmosphere, is assumed to be $\sim l/\sigma_v$. 

The differential X-ray luminosity profile with projected radius for Abell 2029 obtained with the Chandra X-ray Observatory is shown in Figure 6 (black curve).  Superposed on the black curve are two models based on Equations 1, 2, and 3. The luminosity profile can be reproduced by varying $l$ while holding the velocity dispersion to the measured value of $\sigma_v=169~\rm km~s^{-1}$ \citep[][]{XRISM_A029}.  This profile is indicated in red. 

The turbulent power profile is able to match the luminosity profile assuming a constant turbulent speed if $l$ rises with radius from a value of 0.5 kpc in the inner two kpc to roughly $20$ kpc where the profiles cross the cooling region at about 120 kpc.   A justification for an injection scale rising with radius is discussed below.
\begin{figure}
    \centering
    
    \includegraphics[width=\columnwidth]{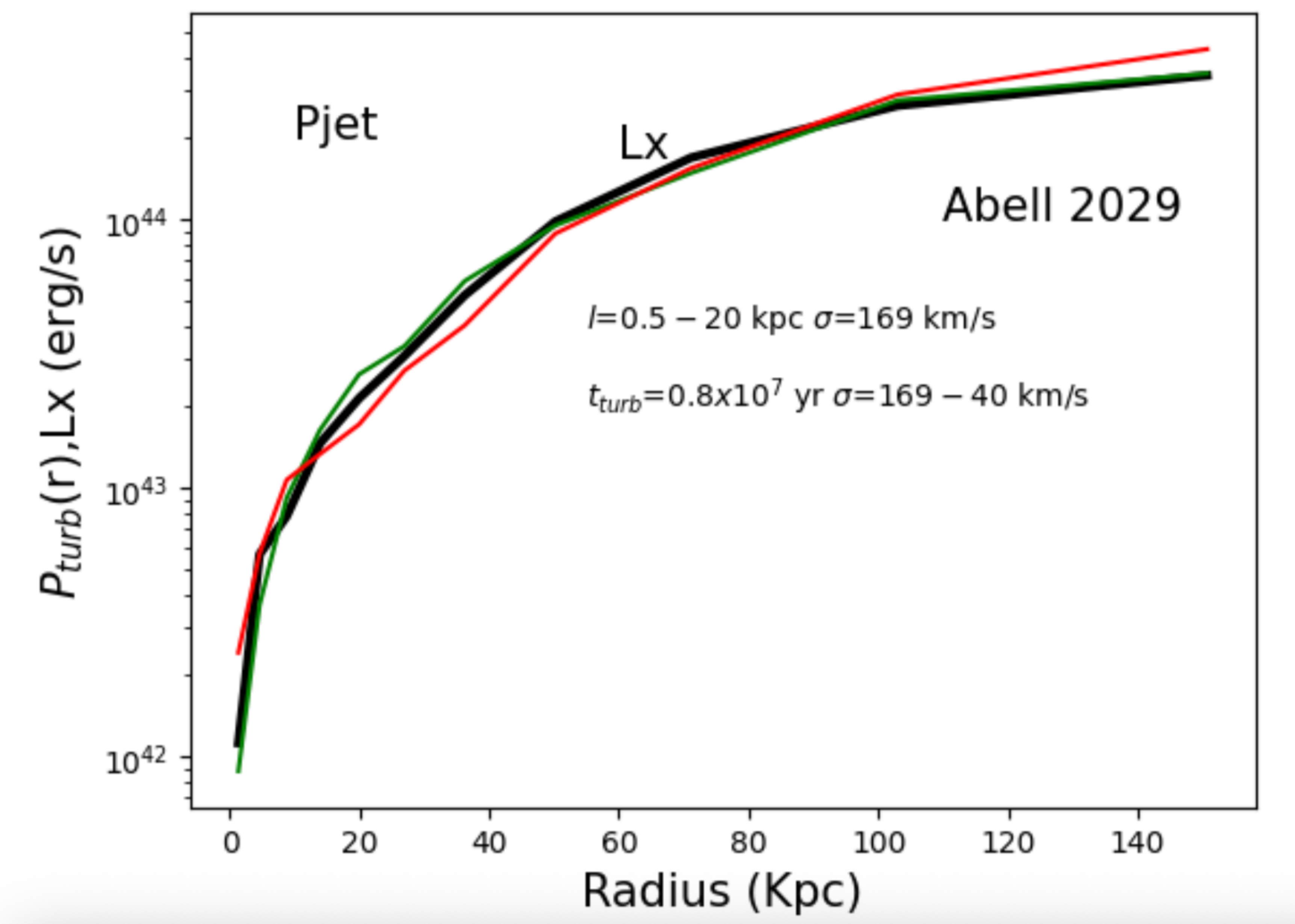}
    \caption{Phenomenological heating models vs cooling in black. The red curve assumes a constant $\sigma_{\rm turb}=169 ~\rm km~s^{-1}$ with a radially-rising injection scale.  The green curve assumes a constant injection scale and constant dissipation timescale and a radially-declining velocity dispersion.  It is unclear either model is physically plausible}
\end{figure}

Conversely, assuming a constant turbulent dissipation timescale $\tau\lesssim 10^7$ yr the luminosity profile can be reproduced (shown in green) with a radially declining velocity dispersion lying between the observed $\sigma_v=169~\rm km~s^{-1}$ and declining to $\sigma_v\sim 40~\rm km~s^{-1}$ in the outer cooling region. This profile implies an injection length scale $l\lesssim 1.7$ kpc over the entire volume which is probably implausible.  We don't know the radial dependence of $\sigma_v$ in the inner regions of Abell 2029.  

This example demonstrates that Equations $1-3$ have arbitrary solutions that may or may not be plausible.  Both solutions are physically challenging.  A short dissipation timescale corresponding to a radially declining velocity dispersion throughout the cooling region would require energy to be injected rapidly on small spatial scales over a large volume. This energy must be replenished in $\sim 10^7~\rm yr$ in the inner $20-40$ kpc and within a few $10^8~\rm yr$ over the cooling volume to maintain turbulent heating in a steady state.  Jets would struggle to achieve this.  Solutions involving radially rising injection scales from jets and bubbles may be plausible if the velocity dispersion remains high throughout the cooling volume where the gas density is falling and the cooling time is lengthening.  

The fits between heating and cooling are aided by similar scaling on the left and right hand sides of Equations 3 and 4. $L_{\rm x}\propto n_e^2 \Lambda$, where $\Lambda$ is the cooling function. On the right side of the equation $M_g/r \sim n_e/r$ and $n_e \sim 1/r$.  Therefore, both sides of the equation scale as $1/r^2$.  Unique fits are clearly not possible and instead must be evaluated by their physical plausibility.  

Similar approaches were taken in the XRISM analyses of Hydra A  \citep[][]{2025Rose}, M87  \citep[][]{2025XRISM_M87}, and Perseus  \citep[][]{2025XRISM_Perseus}.  In these studies the injection scales and dissipation timescales were constrained at low-power using the effective length scale, which under-powers turbulence by factors of several to more than an order of magnitude.  The effective scale is an estimate of the largest scales contributing to $\sigma_v$.  It need not be the energy injection scale required to balance cooling.  Therefore, smaller injection scales of order the bubble sizes and dissipation timescales comparable to the duty cycle were used to boost the power to approximately match the cooling luminosities.  While perhaps plausible, these models are phenomenologically motivated not physically motivated.  


\section{Physically-motivated Models of Turbulent Dissipation}

\subsection{HAK19 Turbulent Dissipation Model}
Using a general form of Equation 2, \citet[][]{HAK19}, referred to hereafter as HAK19, proposed a turbulent heating model that anchors the key timescales to the sound speed $c_s$. Their model assumes the turbulent heating timescale $t_{\rm heat}=\alpha^{-3/2}t_{\rm sound}$.  Here $\alpha$ is the \cite{Shakura_Sunyaev} viscosity parameter which characterizes the ratio of turbulent energy to thermal energy of the atmosphere.  In their model $\alpha$ is a constant between $\sim 5\%$ and $10\%$, which is within the observed range \citep[][]{2025Rose,XRISM_2025_A2029} and Figure 3.  They further assume the turbulent injection scale rises linearly with radius as $l=R$.  This assumption allows turbulence to impart energy onto the gas at large radius because energy generated centrally by jets would be unable to propagate outward.  The model is agnostic to the origin of turbulence.  The HAK19 model reduces the dimensionality of the problem by tying the turbulent velocity dispersion to the sound speed as $\sigma_{\rm turb}\simeq \alpha^{1/2}c_s$. 
Using this approximation, we compare the HAK19 model directly to both the XRISM and Chandra data.   



The HAK19 model for Abell 2029, shown in Figure 7, is able to offset radiation losses throughout the cooling region with $\alpha=0.12$.  The model predicts a radially-rising velocity dispersion lying between $\sigma =156-298~\rm km~s^{-1}$ within 120 kpc. The observed value, $\sigma \sim 169 ~\rm km ~s^{-1}$ within 180 kpc \citep[][]{XRISM_A029}, is broadly consistent with the range of model predictions.  However, the observed velocity dispersion lies well below the extremes at large radii. 
Figure 4 is generally inconsistent with radially-rising velocity dispersions, which is a feature of the HAK19 model. Radially-resolved velocity profiles are generally flat or declining as observed in Hydra A, Perseus, and M87. 

\begin{figure}
    \centering
    
    \includegraphics[width=\columnwidth]{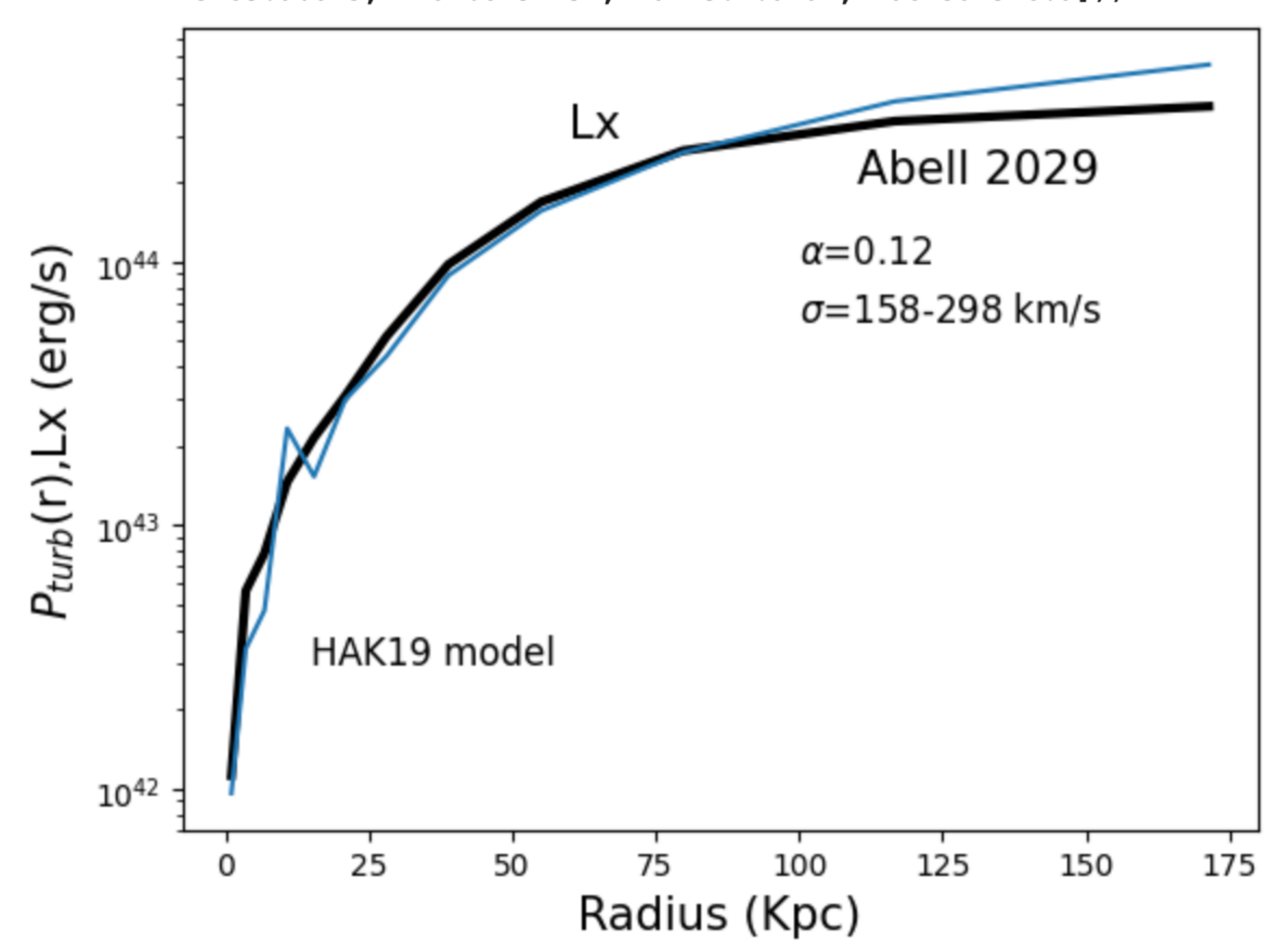}
    \caption{HAK19 heating model is shown to balance cooling over the entire cooling region. The model features a radially-rising injection scale with a constant ratio of kinetic to thermal energy $\alpha=0.12$. The model predicts a radially-rising velocity dispersion which is generally not observed.}
\end{figure}

\subsection{Buoyancy-Driven Turbulent Dissipation Model}

While constraints on velocity dispersion profiles and the ratio of kinetic to thermal energy are in tension with the HAK19 model, solutions with a radially varying injection scale seem unavoidable.  A radially-rising injection scale may be a natural consequence of jet turbulence on small scales and bubble-driven turbulence on larger scales.   

Assuming bubbles effectively generate turbulence as they rise, the injection scales may grow as the bubbles expand in the radially-declining pressure field.  \cite{2008Diehl} examined the sizes of X-ray bubbles as a function of atmospheric altitude and found a linear trend.  Bubble diameters are on average consistent with their altitudes to a projected altitude of 200 kpc.  This trend was found within the cooling regions of clusters and provides the physical motivation for an injection scale increasing with altitude.   

 Turbulence injected by rising bubbles will inject energy roughly at the rate of the bubbles' terminal speeds, $\tau \simeq l/\epsilon c_s$. Here $\epsilon$ is the ratio of the bubble terminal speed to the sound speed $c_s$.  Motivated by \citet[][]{2008Diehl} and assuming $l\propto R$, we arrive at the expression,
\begin{equation}
  P_{\rm buoy} \simeq {3 \over 2} M(r){\sigma_v^2\epsilon c_s \over R}. 
\end{equation}

Here, $\epsilon$ is assumed for the moment to be $\sim 0.5$ \citep[][]{Churazov2001, 2006Rebusco}, and $R$ is the radius or atmospheric altitude.  The value of $\epsilon$ will vary depending on the level of mass loading \citep[][]{Gingras2024}, bubble shape \citep[][]{Churazov2001,Zhang2022}, and perhaps magnetic field stresses \citep[][]{2006Lyutikon, 2008Dursi}.  It cannot approach or exceed unity. This expression is used to model the X-ray luminosity profile by varying $\epsilon$ constrained only by $\sigma_v$ as a function of radius.

Equilibrium between energy injection and radiation at all radii is assumed so that $t_{\rm turb}=t_{\rm rad}$. The injection scale is then $l=\sigma_vR/\epsilon c_s$, which is proportional to the altitude $R$ but foreshortened by the ratio of the turbulent speed to the terminal speed. 

\subsection{Equilibrium between Radiation and Heating}

The ability to maintain equilibrium depends on the jet duty cycle.  Based on systems with multiple bubble pairs, typical duty cycles range between $\sim 6\times 10^6~\rm yr$ for M87 to $\sim 10^8~\rm yr$ for Hydra A \citep[][]{Vantyghem2014}.  This range is broadly consistent with the timescales shown in Figure 5.  

Equilibrium need not be maintained at all times for turbulence to heat effectively. But strong departures from equilibrium will affect the atmosphere's thermodynamic properties.  The ratios of kinetic to thermal energy are only a few percent with small variance (Figure 3).  Therefore, at any moment turbulence provides a small fraction of the atmosphere's thermal energy. A departure such that $t_{\rm turb} \gg t_{\rm buoy}$ would allow energy to accumulate on large scales driving up the turbulent speed $\sigma_v$. Energy would then inefficiently cascade to small scales.  If the system is underpowered so that $t_{\rm turb} \ll t_{\rm buoy}$, turbulence will decay thus reducing the dissipation rate and increasing atmospheric cooling. Increasingly-powerful AGN outbursts would then be required to replenish radiation losses.

 
Atmospheres experiencing dramatic excursions away from equilibrium would yield large variations in $\sigma_v$, which are not seen in  Figures 1 \& 3. The mean velocity dispersion and $1\sigma$ variation found for the jetted atmospheres in Table 1 is $165\pm 60$.  Both the mean and variance are low across nearly four decades of jet power. A much larger sample is needed to sample the full range  of $\sigma_v$ and its variance. 

\begin{figure}
    \centering
  \includegraphics[width=\columnwidth]{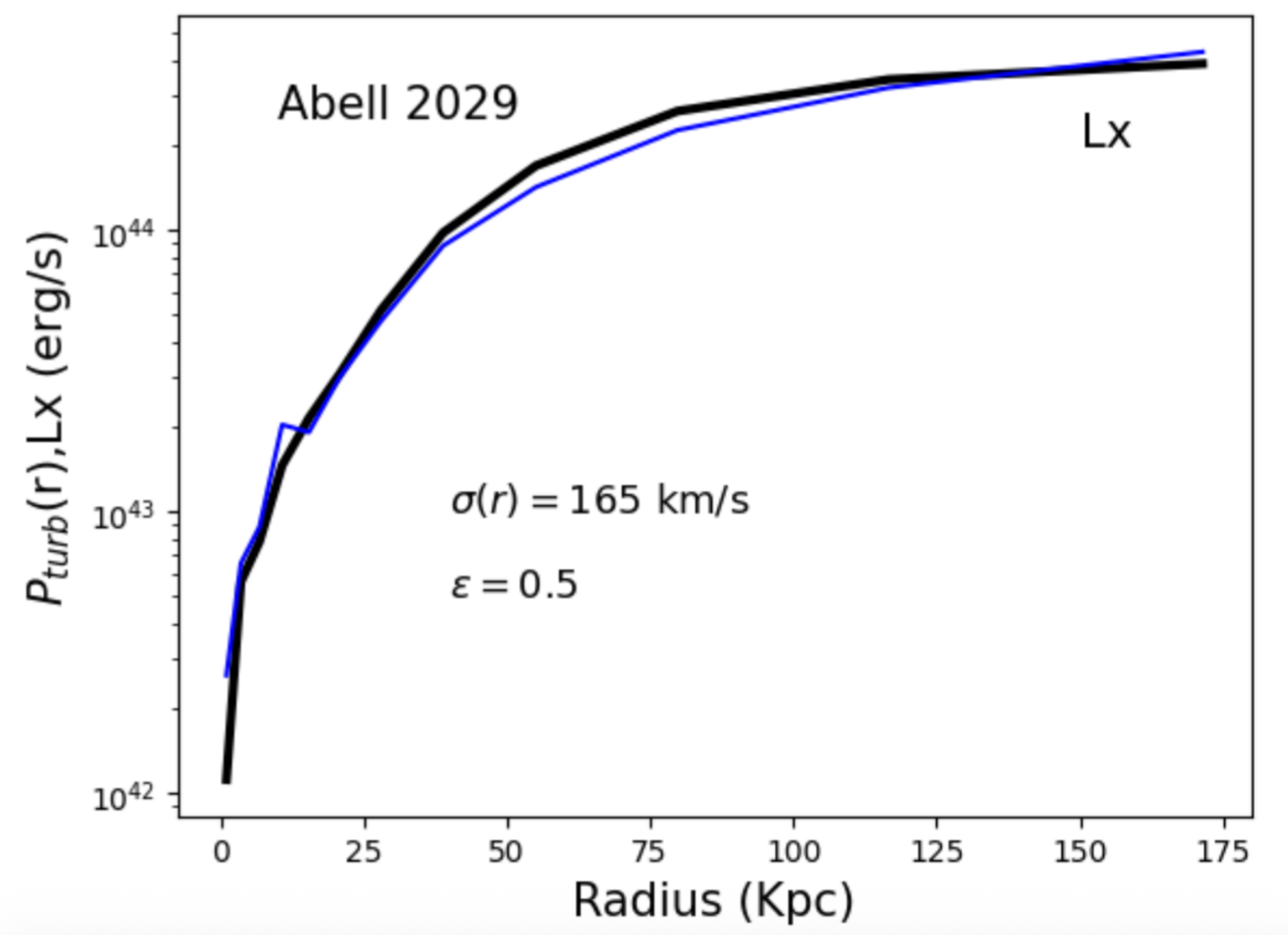}
    \caption{Buoyancy heating model shown in blue is able to offset cooling losses (black) over the entire cooling volume with a bubble speed $\epsilon=0.5$ of the sound speed. }
\end{figure}

\subsection{Testing the Buoyancy Model on Abell 2029}

We test the buoyancy model on Abell 2029 adopting the observed $\sigma_v \simeq 160~\rm km~s^{-1}$ across the cooling region.  The model is able to offset cooling losses over the cooling volume with a buoyancy speed of $\sim 0.5c_s$ (Figure 8).  This is true despite Abell 2029's unusually high cooling luminosity which is twice that of Perseus \citep[][]{Rafferty2006}. 

The relatively high turbulent power in Abell 2029 is due to two factors.  First, its high atmospheric temperature rising from 2.4 keV in the center to 8 keV at the cooling radius implies a higher terminal speed than is achievable in a cooler atmosphere.  For example, at an altitude of 20 kpc Perseus' atmospheric temperature (discussed below) is $\sim 3.6$ keV while Abell 2029's is $\sim 6$ keV. Therefore, Abell 2029's sound speed is approximately 30\% higher than in Perseus. The terminal speed is therefore proportionally larger for its $\sigma_v$ and $\epsilon$ leading to higher turbulent power. Second, a constant velocity dispersion with radius is conducive to higher turbulent power at high altitudes. 

We attempted to resolve $\sigma_v$ into two regions within the cooling volume using spatial-spectral mixing techniques, but failed to find a significant gradient.  If a future, more sensitive, observation of Abell 2029 finds a declining velocity dispersion within the cooling radius, $\epsilon$ must rise and the injection scales must decline to generate enough power to offset cooling. This positive result for Abell 2029 should be taken with caution.


\subsection{Hydra A: Powerful Jets With a Resolved \texorpdfstring{$\sigma_v$}{sigma-v}}

XRISM observed the central $\sim 100$ kpc of the Hydra A cluster in 2024 November.  An analysis of the full footprint centered on its powerful radio source and cavity systems \citep[][]{Wise2007} yielded a central velocity dispersion of $164 \pm 10 ~\rm km~s^{-1}$ \citep[][]{2025Rose}. A measurement of the turbulent energy density and an analysis of the plausible range of turbulent power concluded that turbulence would struggle to offset atmospheric cooling but could not be ruled out.  

A second 200 ks observation was obtained in 2025 May toward the cluster-scale cavity 190 kpc north of the nucleus. The Northern cavity contains much of the energy expended by the radio source over the past 100 Myr.  A subarray spatial analysis of both pointings is presented in Majumder et al. (2026, in preparation). The analysis reveals significant structure in the velocity dispersion field within $\sim 200$ kpc.  The mean dispersion drops modestly from $\sigma_v=160 \pm 35 ~\rm km~s^{-1}$ in the inner 12 pixels to $\sigma_v=130 \pm 55 ~\rm km~s^{-1}$ beyond to the roughly 100 kpc edge of the cooling region.  The northern pointing yielded a mean $\sigma_v=140 \pm 25 ~\rm km~s^{-1}$.  Therefore, Hydra A's $\sigma_v$ profile is nearly flat from the centre out to roughly 280 kpc. 

Turbulence generated by the bubbles may maintain the velocity dispersion over the full extent of the cavity system.  However, the effective length scales are much larger toward the outer cavities than in the core, making $\sigma_v$ harder to interpret. Absent a pointing away from the cavity system to the East or West, we cannot be sure $\sigma_v$ primarily represents jetted turbulence. The central two dispersions here are adopted to model the cooling region using the buoyancy prescription and to compare to the HAK19 model.  A detailed description of the data analysis is given in an accompanying paper (Majumder et al. 2026, in preparation).

\subsection{HAK19 Model Applied  to Hydra A}

A HAK19 model fit to Hydra A's differential luminosity profile is shown in Figure $9$.  A value of $\alpha=0.15$ provides a good fit to the luminosity profile throughout the cooling volume.  The model appears to overshoot the luminosity profile beyond 60 kpc.  However, the luminosity dip is due to the giant X-ray cavities at those altitudes. As in Abell 2029, the one parameter HAK19 model recovers the luminosity profile remarkably well.   

The HAK19 model predicts a one-dimensional velocity dispersion profile of $\sigma_v=175 ~\rm km~s^{-1}$ in the center rising to $\sigma_v=230~\rm km~s^{-1}$ at about 120 kpc.  The predicted central $\sigma_v$ is consistent with the upper error bar of the observation.  However, the HAK19 $\sigma_v$ at larger radii exceeds the observed value by $\sim 100~\rm km~s^{-1}$ thus is inconsistent with the XRISM measurement.  Furthermore, the value of $\alpha$ required to match the luminosity profile exceeds the observed ratio of kinetic to thermal energy in Hydra A \citep[][]{2025Rose} by more than a factor of two. 
The rising velocity dispersion inherent to HAK19 models with a single $\alpha$ parameter is inconsistent with the flat or radially declining velocity dispersion profiles in most systems.  This analysis highlights the powerful constraints provided by velocity field measurements within the cooling regions of cluster atmospheres. 
\begin{figure}
    \centering
    
    \includegraphics[width=\columnwidth]{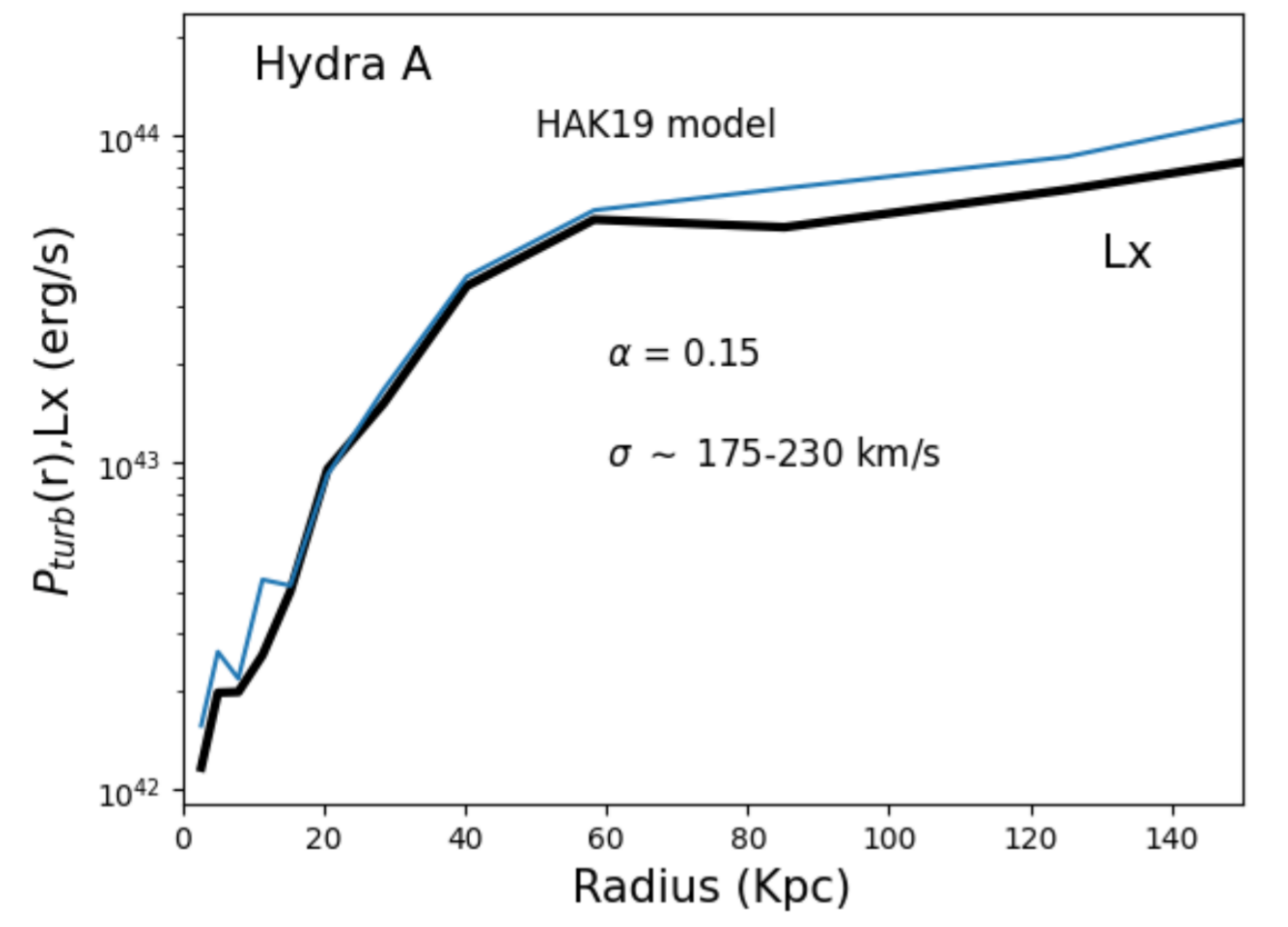}
    \caption{HAK19 heating model fit with a radially-rising injection scale and constant ratio of kinetic to thermal energy $\alpha=0.15$. While the model is able to offset cooling throughout the cooling volume, the model predicts a radially-rising velocity dispersion which is not observed.  }
\end{figure}

\subsection{Buoyancy Model Applied to Hydra A}

The buoyancy model applied to Hydra A constrained by two XRISM velocity dispersion measurements is shown in Figure 10.  Adopting $\epsilon =0.2$ reproduces the X-ray luminosity very well within 60 kpc.  To fit the outer volumes of the cooling region $\epsilon$ must rise to about $0.5$. The rise in $\epsilon$ responds to the slightly declining velocity dispersion.  However, the error in the dispersion is large.  Adopting a milder dispersion drop would lower $\epsilon$ to a value of $0.2-0.4$ at altitudes beyond 60 kpc, again showing the power of a resolved velocity field to constrain the model.  

A larger $\epsilon$ implies faster bubble speeds, smaller injection scales, and higher turbulent power near the powerful outer bubbles. 
\begin{figure}
    \centering
  \includegraphics[width=\columnwidth]{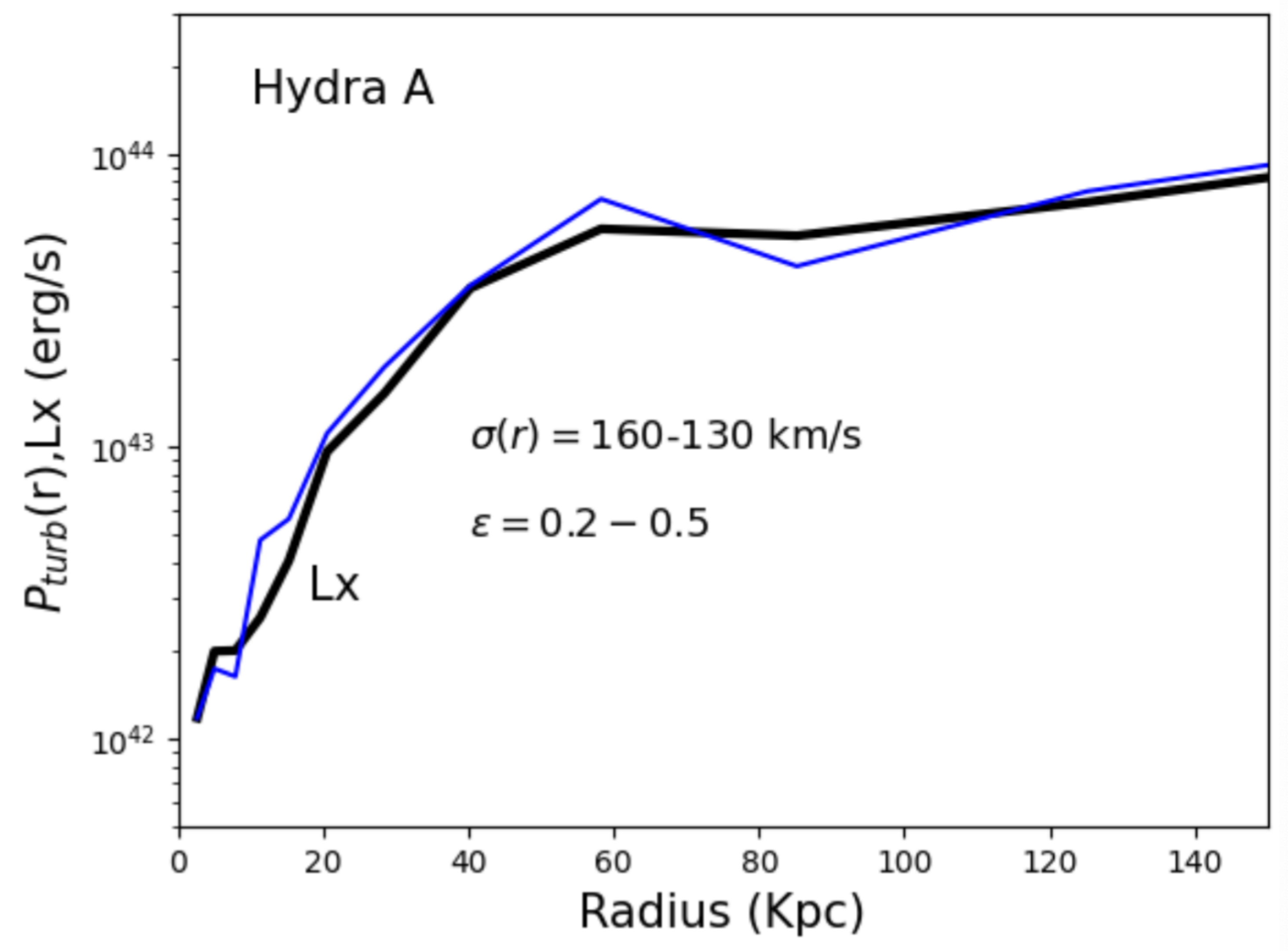}
    \caption{Buoyancy heating model applied to Hydra A (blue) is able to offset cooling losses (black) over the entire cooling volume with a bubble speed $\epsilon=0.2-0.5$ times the sound speed. }
\end{figure}
The lower value of $\epsilon \lesssim 0.5$ would be consistent with mass loading by metal-enriched gas being lifted behind the bubbles \citep[][]{2008Simionescu,Kirkpatrick2009,2011Gitti}.  The values of $\epsilon$ required to fit Hydra A are physically plausible.

\subsection{Comparing the Hydra A Buoyancy Model to Simulations}

\cite{2012Vazza_Roediger} considered turbulence using a magneto-hydrodynamic (MHD) simulation tailored to the jet power of Hydra A. They were able to distinguish between projected large-scale turbulence and centrally-concentrated jet and bubble turbulence. \citet{2012Vazza_Roediger} found turbulent rolls $\sim 10-20 ~\rm kpc$ in size with turbulent speeds between $200-300\rm ~km~s^{-1}$.  These speeds  correspond to one dimensional velocity dispersions of $115-173\rm ~km~s^{-1}$, which lie within range of the observed values but exceeding them at the high end by $100 ~\rm km~s^{-1}$. The simulated jetted turbulence is patchy with long diffusion timescales. The injection lengths implied by the buoyancy model range between $\simeq 3-20~\rm kpc$ within 40 kpc and  $\simeq 20-30~\rm kpc$ near the 120 kpc cooling radius (see Figure 11). The buoyancy model and simulation agree remarkably well. 


\subsection{Buoyancy Model Applied to the Virgo \& Perseus Clusters}

The analysis presented here and those for M87 \citep[][]{2025XRISM_M87}, Hydra A \citep[][]{2025Rose}, and Perseus \citep[][]{2025XRISM_Perseus} reach broadly similar conclusions. Much of the acceptable solution space in each system lies below the level of heating required to offset cooling.  Only the most optimistic assumptions lead to thermal balance.  For example, equilibrium solutions for M87 require turbulence injected and cascading to small scales on the 12 Myr duty cycle of the jet-driven shocks. The injection scale would then lie below one kpc, which is much smaller than the X-ray bubbles and shock fronts. Whether this solution is sustainable is unclear. 

The buoyancy model applied to M87 is in blue in Figure 12 and the cooling X-ray luminosity in black.  Models are shown for the plausible bubble speed $\epsilon =0.5$ which lies below the cooling profile, and $\epsilon =1.2$ which would be required at minimum to balance radiation losses.  The X-ray luminosity profile is identical to \citet[][]{2025XRISM_M87}.  The two disjoint dashed lines within 5 kpc represent the two bubble speeds assuming the inner $\sigma_v=262~\rm km~s^{-1}$ represents pure turbulence.  It was pointed out earlier that jetted bulk motion cannot be disentangled from pure turbulence within 5 kpc \citep[][]{2025XRISM_M87}.  The model in Figure 12 shows that turbulence at this level would strongly overheat the inner region which would be inconsistent with its thermodynamic profiles.  The buoyancy model is unable to balance radiation losses throughout the cooling volume for a plausible range of terminal speeds $\epsilon <1$.  

Heating the cooling volume of Perseus is similarly problematical.  \cite{2025XRISM_Perseus} found that turbulent injection scales comparable to the effective length scale leads to a turbulent heating rate lying far below the cooling rate over most of the $\sim 100 ~\rm kpc$ cooling volume.  Injection scales much smaller than the effective length scales are required to achieve balance over the cooling volume.  

A fit to the Perseus cluster atmosphere with the buoyancy model is shown in Figure 13.  The X-ray luminosity $L_X \propto \rho^2 \Lambda$ and sound speeds are based on the density and temperature profiles from \citet[][]{2017Tang}. We assume a cooling function $\Lambda = 2.5\times 10^{23} ~\rm erg~cm^{-3}~s^{-1}$ and other parameters described in \citet[][]{2025XRISM_Perseus} to enable a direct comparison to that paper. The cooling region plotted in Figure 13 has been resolved into three largely independent regions: $<20$ kpc $\sigma_v\simeq 171 ~\rm km~s^{-1}$, $20-60$ kpc $\sigma_v\simeq 136 ~\rm km~s^{-1}$, and $60-100~\rm kpc$ $\sigma_v\simeq 71 ~\rm km~s^{-1}$.  The errors on these measurements are approximately $20~\rm km~s^{-1}$ (\citet[][]{2025XRISM_Perseus}.  

Figure 13 shows that the buoyancy model would offset cooling in the inner $20$ kpc for plausible values of $\epsilon \lesssim 0.5$ where the bubbles occupy a significant fraction of the volume. The turbulent injection scales in the inner 20 kpc lie between $3-12 ~\rm kpc$ which are comparable to the sizes of the central bubbles (see Figure 11).  However, beyond $\sim 20$ kpc the turbulent power falls below the cooling luminosity for reasonable values of $\epsilon$.  Assuming $\epsilon=0.5$, the turbulent injection scales beyond 20 kpc lie between $5-10 ~\rm kpc$. Achieving balance between heating and cooling beyond the inner region would require $\epsilon \sim 1-3$, which is implausible.

Bubbles injecting energy at a reasonable fraction of the sound speed would require injection scales of only a few kpc because of the atmosphere's declining radial velocity dispersion and relatively cool temperature.  Injection scales of this size are much smaller than the bubble sizes and vastly smaller than the cooling volume. Whether turbulence with these properties can be generated throughout the atmosphere is unclear.

\begin{figure}
    \centering
    
    \includegraphics[width=\columnwidth]{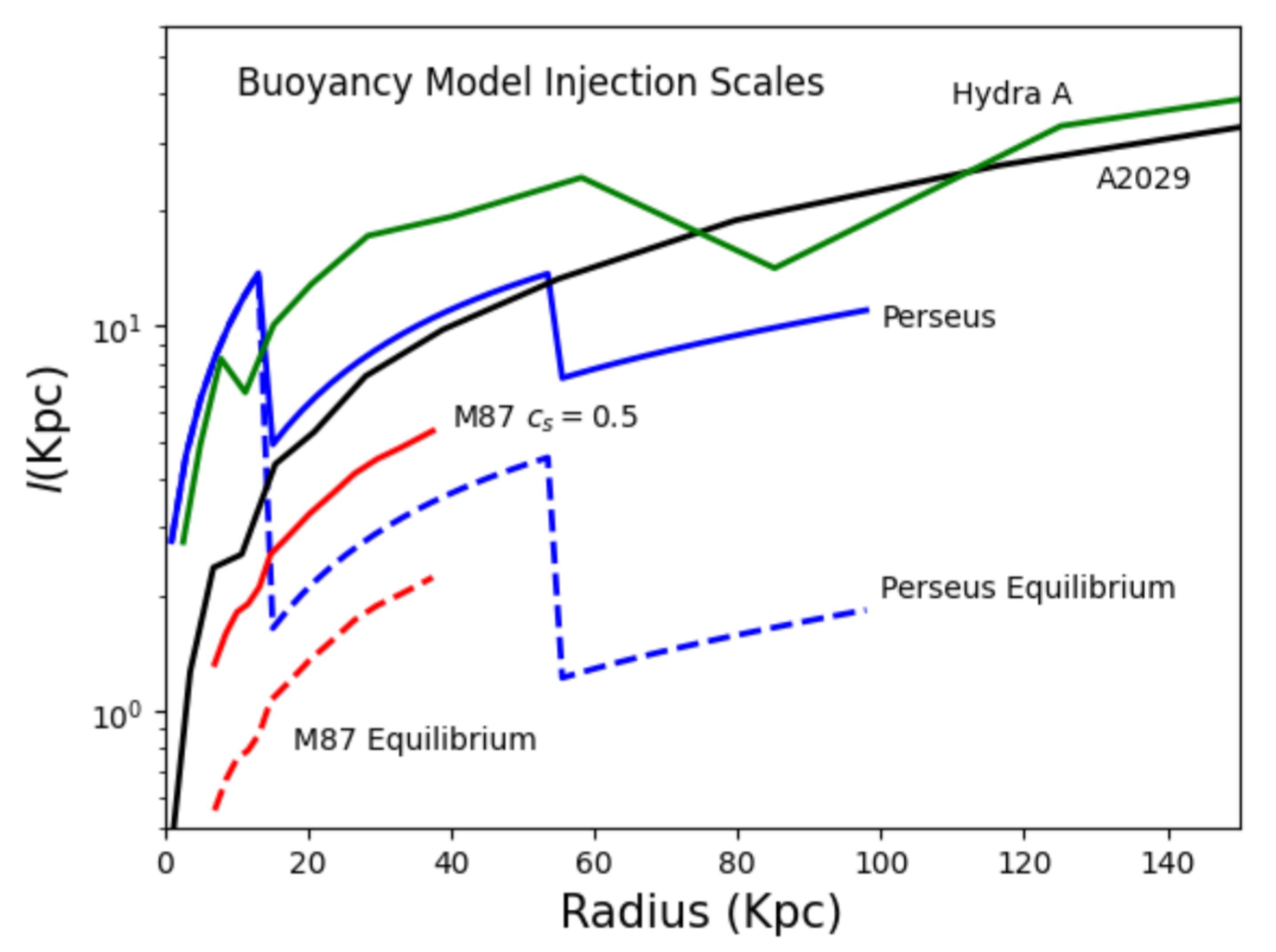}
   \caption{ Injection scale vs radius or altitude for buoyancy models discussed in Section 4.  The abrupt jags in M87 and Perseus are due to $\sigma_v$ binning.  The injection scales for the Perseus and M87 equilibrium models (dashed lines) would require injection faster than the sound speed and small injection scales to balance radiation losses.  }
\end{figure}

\subsection{Comparing the Perseus Buoyancy Model to Simulations}

Simulations tailored to Perseus by \cite{2025Li_Yang} are broadly consistent with our buoyancy model that under-powers cooling in Perseus. However, they conclude heating would be more problematical in the core than the outskirts, perhaps due to the very short duty cycle required to replenish turbulence where cooling is fastest.  The buoyancy model assumes fine balance between heating and cooling which requires turbulence to be replenished at the center in less than $10^8~\rm yr$ due to rapid radiation losses (Figure 5).  The \citet[][]{2025Li_Yang} simulation traces one outburst over a duration of about 130 Myr, which is much longer than the timescales required to maintain fine balance at the center of the atmosphere. 

\begin{figure}
    \centering
  \includegraphics[width=\columnwidth]{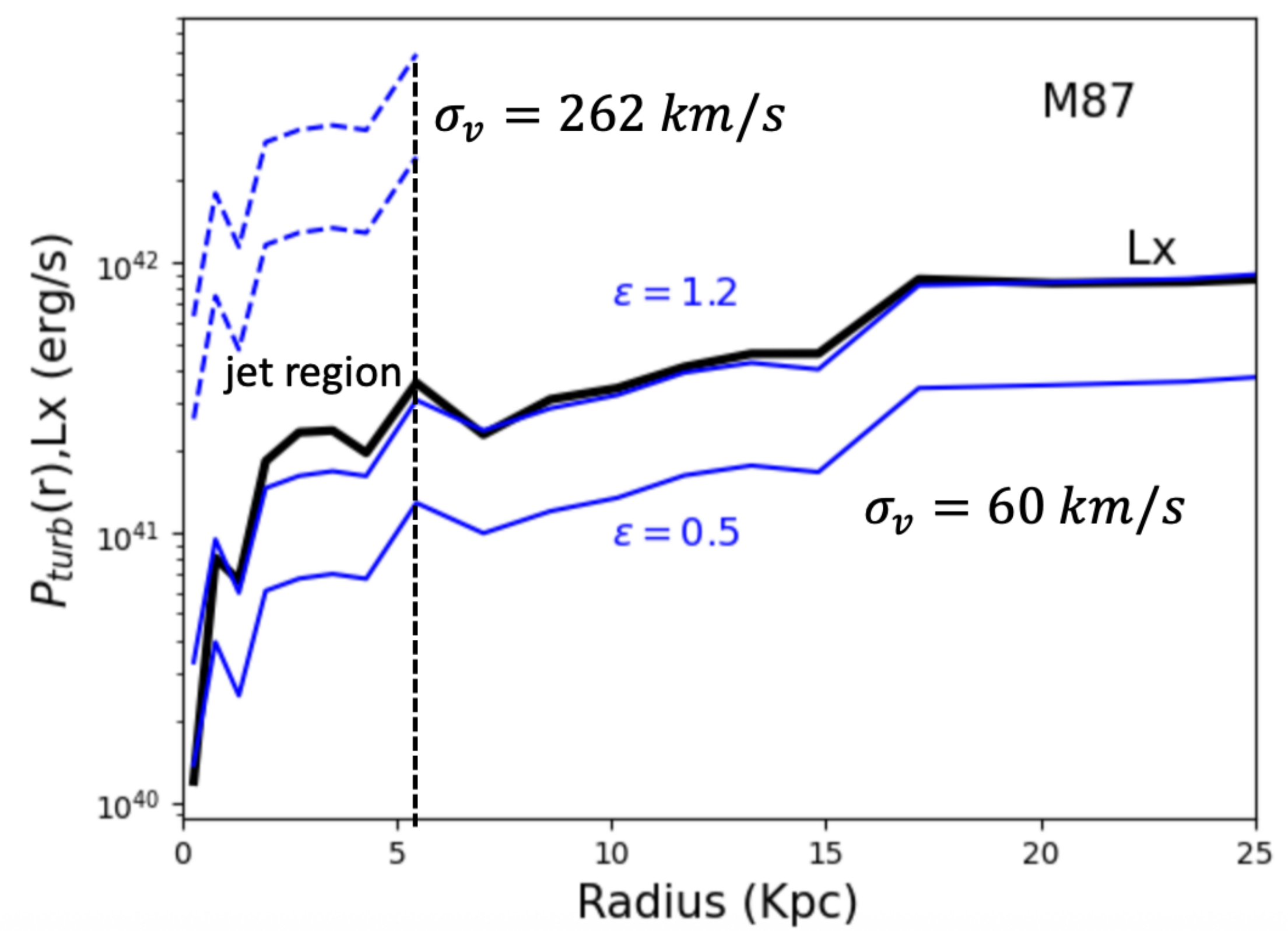}
    \caption{Buoyancy heating model applied to M87 (blue) would require transonic bubble speeds to offset cooling losses (black) over the volume apart from the inner $\simeq 5$ kpc where the jet is likely driving high bulk velocities.  The lower dashed line corresponds to $\epsilon=0.5$ and the upper dashed line corresponds to $\epsilon=1.2$.   }
\end{figure}


The assumption that energy is injected at the rate it is dissipated is unlikely to hold in detail.  Departures from equilibrium would allow cooling to fuel molecular clouds and star formation in Perseus-like clusters \citep[][]{McDonald2018}. However, departures cannot be large or lengthy without upsetting the relatively homogeneous thermodynamic profiles observed in cooling atmospheres \citep[][]{Hogan2017}. 


\begin{figure}
    \centering
    
    \includegraphics[width=\columnwidth]{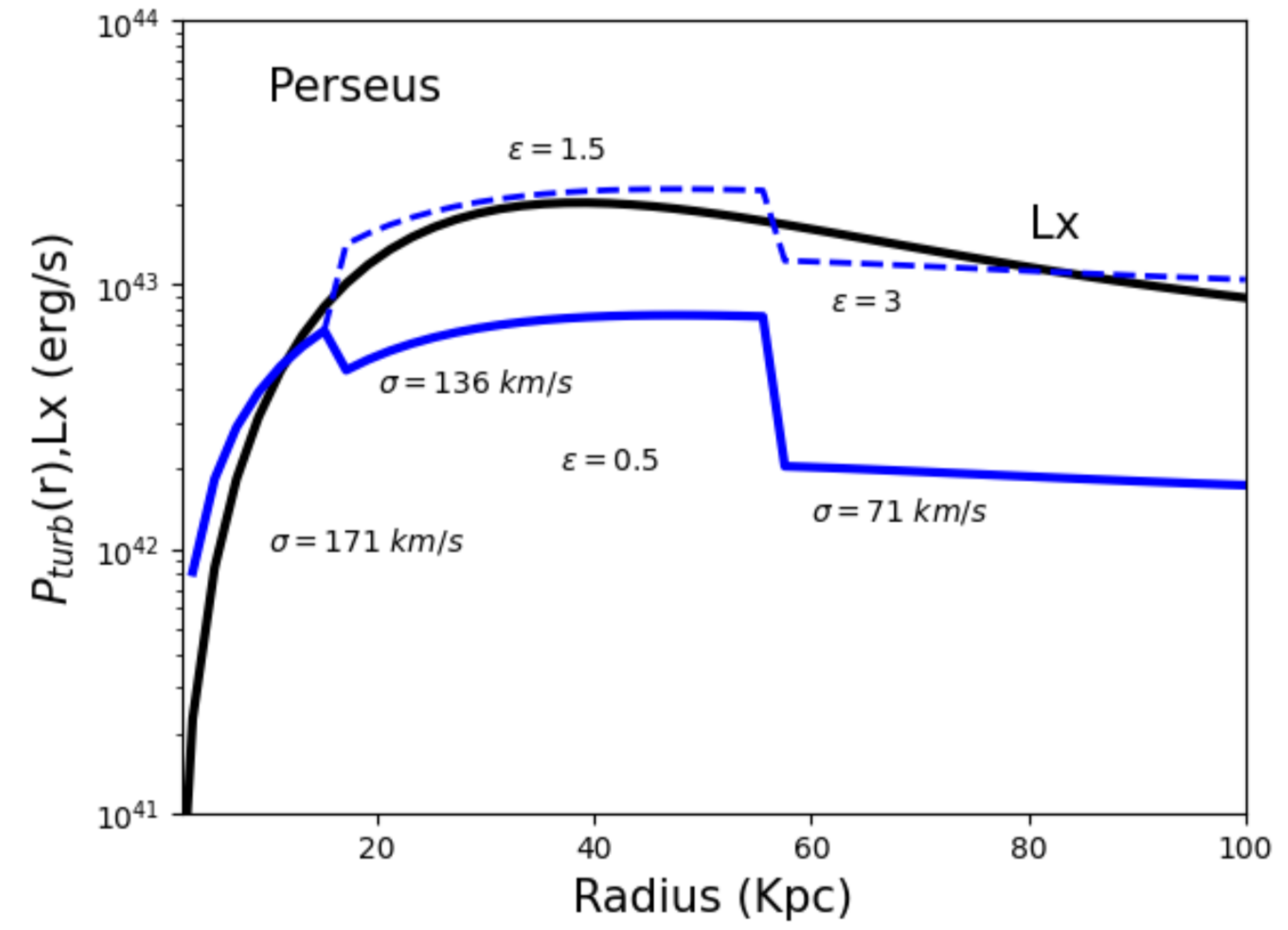}
   \caption{Cooling (black) vs buoyancy heating model (blue) for the Perseus cluster.  The cooling profile is based on the density profile from Tang \& Churazov (2017). The equilibrium model (blue-dashed) requires $\epsilon$ exceed unity beyond about 15 kpc, which is unphysical. Moderate values of $\epsilon$ cannot offset cooling throughout the cooling volume. }
\end{figure}

The unknown fraction of $\sigma_v$ in turbulence local to the cooling volume adds another layer of uncertainty. Models tailored to Perseus \citep[][]{2025XRISM_Perseus,2025Bellomi} found 10 percent or less of the central $\sigma_v$ in projected large-scale motion.  They attribute the central $\sigma_v$ rise to its jets and bubbles. This interpretation is in tension with other hydrodynamic \citep[][]{2025Li_Yang,Bourne2017} and MHD models \citep[][]{2021Ehlert} that found  $\sigma_v$ is primarily transient, jetted bulk flows \citep[][]{2024Truong} or preexisting large-scale turbulence. These models are qualitatively consistent with Figures 1 \& 2 which show no trends between $\sigma_v$ with either radius or jet power. 

\subsection{Turbulent Dissipation Timescales}

The turbulent dissipation timescales implied by Equation 4 are shown in Figure 14.  This figure accompanies the closely-related injection scales in Figure 11.  The timescales fall below $10^7$ yr in the innermost regions of M87 and Abell 2029.  They rise to a few $10^7~\rm yr$ between a few and 20 kpc and rise above $\sim 10^8 ~\rm yr$ toward the outer cooling volumes at $\sim 100~\rm kpc$.  XRISM cannot resolve or constrain the innermost regions where the timescales fall below $10^7~\rm yr$.   However, figures 11 and 14 lead to the unavoidable conclusion that equilibrium cannot be achieved with a single Kolmogorov-like injection scale imparted by the jets and bubbles.  

The timescales in Perseus and Virgo/M87 are shown for the transonic equilibrium models (dotted) and sub-sonic terminal speeds (solid). In order to maintain equilibrium the dissipation timescales should be comparable to the jet/bubble duty cycles \citep[][]{Vantyghem2014,Dunn2006}, which generally lie between $\lesssim 10^7~\rm yr$ in M87 to $\sim 10^8~\rm yr$ in Hydra A.  The duty cycles and equilibrium dissipation timescales generally lie within a factor of a few of each other, indicating perhaps that equilibrium could be achieved in Hydra A.  

However, achieving equilibrium in the outskirts of Perseus and M87 would require very short duty cycles with turbulent energy transported transonically. Such fast energy injection would be infeasible by bubbles but not by shocks or sound waves.  Their injection scales being only a few kpc or less in the inner regions are correspondingly small.  While sub-sonic bubble transport appears to be infeasible, mechanisms that can inject turbulence transonically are preferred. 

\subsection{Stratification}

In a stratified atmosphere, gravity can prevent turbulent eddies from turning-over radially \citep[][]{1990sidi,2020Mohapatra} which is concerning for the buoyancy model.   Gravity limits the size of a radial eddy to be smaller than the ratio of $\sigma_v$ to the \BV{} frequency (see Appendix Equation A4). 
Equation A4 leads to a lower limit on the turbulent dissipation timescale given by Equation A8.  This in turn depends on the radial derivative of the entropy index $\eta = d\ln K/d\ln R$ and the Kepler speed $v_K$.  

Both were calculated for Hydra A to check the model. We found $\eta \sim 0.75$ leading to a buoyancy terminal speed  $\epsilon c_s \simeq 2/3v_K$.   The dissipation timescale based on Equation A8 is the purple dashed line in Figure 14.  This line runs along and well within a factor of two of the buoyancy model dissipation timescale for Hydra A shown in green.  Therefore, radial eddy modes with sizes $l$ implied by Figure 11 are physically plausible. 

\begin{figure}
    \centering
  \includegraphics[width=\columnwidth]{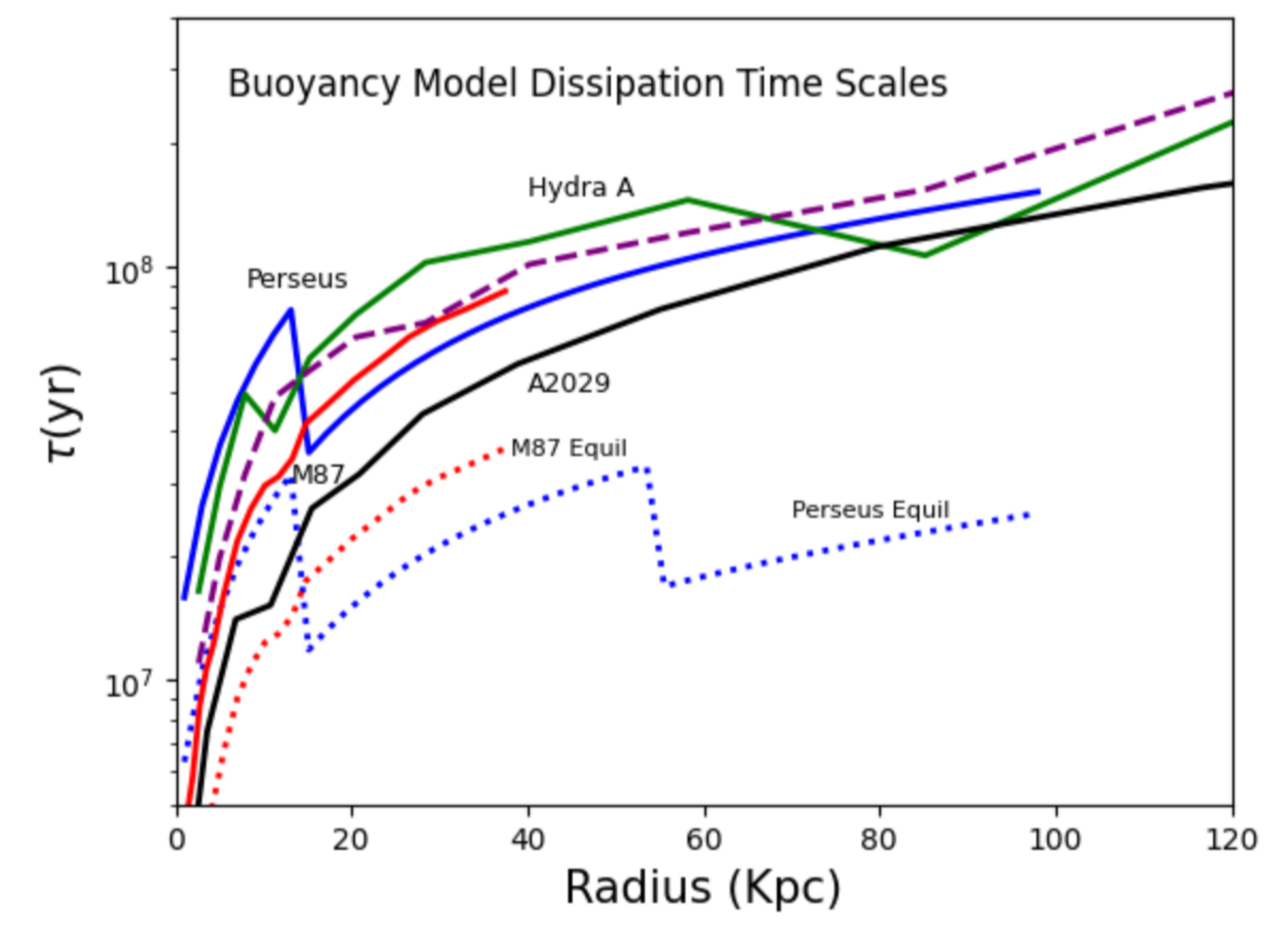}
    \caption{Buoyancy heating model dissipation timescales. The purple dashed line is the Hydra A dissipation timescale based on Equation 12.  Its close agreement to the buoyancy model dissipation timescale shown in green indicates that isotropic turbulence is probably a valid model. The blue and red dotted lines are the Perseus and M87 equilibrium models, respectively. The blue and red solid lines are the Perseus and M87 models with non-equilibrium, sub-sonic terminal bubble speeds.   }
\end{figure}

\section{Summary}

No trend is found between radio jet power and cluster atmospheric velocity dispersion over nearly four decades of jet power. Central velocity dispersions lie between approximately $50~ \rm km~s^{-1}$ and $250 ~\rm km~s^{-1}$.  Their mean is  $<\sigma_v> = 165\pm 60 ~\rm km~s^{-1}$ over a jet power range $\sim 10^{42}-10^{46}~\rm erg~s^{-1}$.  No trends are found between jet power and the ratio of kinetic to thermal energy, or jet power and energy per unit gas mass. To the degree that $\sigma_v$ can be attributed to jets and bubbles, they apparently heat gently by imparting a roughly constant kinetic energy per gas mass.   

Central velocity dispersions in Hydra A \citep[][]{2025Rose}, Perseus \citep[][]{2025XRISM_Perseus}, and M87 \citep[][]{2025XRISM_M87} are consistent with less than $20-50\%$ of their jet power released in turbulemt motion. 
These upper limits do not account for unrelated projected motion along the sight-line or unresolved bulk motion driven by the jets and bubbles which may not thermalize promptly.

The inward-rising velocity dispersions found in Virgo/M87 and Perseus may be jet-powered motions, but with an unknown fraction of turbulence.  For example, the central velocity dispersions in M87 and Cygnus A \citep[][]{2025Majumder} are probably largely jet-driven bulk motions.  

A heating model that assumes turbulence is injected at the terminal speeds of rising bubbles was presented and applied to several systems.  The model is ideal: $\sigma_v$ is assumed to represent pure turbulence within the cooling volume injected in equilibrium with radiative cooling and turbulent dissipation via a Kolmogorov cascade. 
 
 Abell 2029 was modeled successfully with the buoyancy prescription (Equation 4) for bubbles rising at or below half the sound speed.  The model's success is aided by a single $\sigma_v$ measurement across the cooling volume and by its high sound speed. 
 
 Hydra A was successfully modeled with bubbles moving at roughly half the sound speed aided by its flat $\sigma_v$ profile out to and beyond the cooling volume. It's cluster-scale cavity system fills the XRISM footprint along the jet axis.  Whether heating can be sustained in equilibrium throughout the cooling volume is unclear.  
 
 The buoyancy model applied to Perseus and Virgo/M87 failed beyond the central regions primarily because of their radially-declining velocity dispersion profiles. Equilibrium cannot be sustained by turbulent heating in their outer cooling volumes without injection scales of only a few kpc and short replenishment timescales that cannot be achieved with subsonic turbulent injection.


\subsubsection{Problems and Caveats}

\begin{itemize}

\item Turbulent injection scales lying between a few to $\sim 30$ kpc and dissipation timescales lying between $\lesssim 10^7~\rm yr$ to roughly $10^8~\rm yr$ are required to offset cooling over the cooling volumes (Figures 11 \& 14).  Turbulence injected below the sound speed cannot meet these requirements in Perseus and Virgo/M87.  

\item Neither the fraction of $\sigma_v$ in gaussian random velocity fields or the shape of the turbulent velocity spectrum are known.  Jets and bubbles probably produce bulk motion through momentum exchange as they advance and expand.  Resolution and effective scale-length issues limit our ability to disentangle turbulent motion in the cooling volume from preexisting large-scale motion that will thermalize on timescales exceeding the cooling time \citep[][]{Bourne2017,2012Vazza_Roediger, 2021Ehlert,2025Li_Yang}.  The velocity spectrum must be close to a Kolmogorov power law to heat effectively.

\item Turbulence generated by jets and bubbles must be injected locally because it cannot diffuse throughout the cooling volume.  Apart from the inner cooling regions much of the cooling volume is free of bubbles and will be heated with difficulty by jetted turbulence.  Redirected jets and/or zig-zagging bubble trajectories may be able to generate turbulence on larger scales \citep[][]{2006Heinz, vernaleo2006,2006Dunn2,2013Babul,2018Cielo}.



\item Heating and cooling equilibrium assumed here need not be strictly maintained to offset cooling effectively.  However, large departures would lead to a buildup of power at large injection scales (small wave numbers) which would thermalize too slowly. Conversely, if power is injected infrequently, it will be radiated away before it can heat the cooling volume.



\item The combination of low atmospheric velocity dispersions, anisotropic jet and bubble trajectories, and long turbulent diffusion timescales are severe challenges facing jetted turbulence heating models.  The prospects for turbulent dissipation as a primary or significant heat source in cooling atmospheres are not promising.  A larger sample of cluster atmospheres is required to reach a definitive conclusion.
\end{itemize}


\begin{acknowledgments}
BRM acknowledges the Canadian Space Agency, the Natural Sciences and Research Council, and the Waterloo Centre for Astrophysics for generous financial support.  BRM thanks Niayesh Afshordi, John Zuhone, Mark Voit, Chris Reynolds, and Karen Yang for their insights.  
\end{acknowledgments}

\begin{contribution}
The paper grew from conversations about turbulence between BRM, ACF, HRR, AS, and PEJN. All contributed to the intellectual development of the paper. AM analyzed the Hydra A data and commented on the draft.  PEJN wrote the Appendix. EDM and AS contributed the Abell 1795 data and commented on the draft.  BRM performed the model analysis, developed the buoyancy model, and wrote the paper.   
\end{contribution}
\facilities{Chandra(ACIS), XRISM(Resolve)}

\software{ HEASoft, Xspec \& PyXspec,
          Astropy,
          Matplotlib,
          NumPy,
          Python,
          CIAO}

\appendix
\section{Turbulent Dissipation in a Stably Stratified Atmosphere}
\subsection{Size of Radial Eddies}

In a stably stratified atmosphere, gravity imposes an upper limit on
the radial size of eddies.  Stable stratification requires a positive
radial entropy gradient, so that, when an element of fluid is raised
or lowered, its entropy is, respectively, lower or higher than that of
its new surroundings \citep[][]{1990sidi,2020Mohapatra}  The net force on the fluid element due to
buoyancy and gravity is then directed back towards its equilibrium
position.  In the linear approximation, the equation of motion for
small radial displacements of a fluid element is
\begin{equation}
  \frac{d^2}{dt^2} \deltar = - \ombv^2 \deltar,
\end{equation}
for a displacement $\deltar$, where the \BV{} frequency is given by
\begin{equation}  \label{eqn:ombv}
  \ombv^2 = \frac{g}{\gamma} \frac{d\ln K}{dR}
  = \frac{\eta}{\gamma} \frac{\vkepler^2}{R^2},
\end{equation}
$\gamma = 5/3$ is the ratio of specific heats, $K = kT /
\nelec^{\gamma - 1}$ is the entropy index, $g = \vkepler^2 / R$ is the
acceleration due to gravity, $\vkepler$ is the Kepler speed and $\eta
= d\ln K/d\ln R$.

An eddy turning over radially is subject to this same net restoring
force, so that the maximum radial distance it can reach from its
equilibrium position, $\drmax$, is related to its maximum radial
speed, $\vrmax$, in the usual manner for simple harmonic motion,
\begin{equation}  \label{eqn:vrmax}
  \vrmax = \ombv \drmax.
\end{equation}
For the clusters observed with XRISM, $\vrmax$ cannot significantly
exceed the turbulent velocity dispersion, $\sigv$, so that the radial
extent of the turbulent eddies will be limited to
\begin{equation}  \label{eqn:leddy}
  \leddy \lesssim \frac{\sigv}{\ombv}
  = \sqrt{\frac{\gamma}{\eta}} \frac{\sigv}{\vkepler} R.
\end{equation}
Turbulent eddies larger than this may still turn over transverse to
the radius, but turbulence on larger scales has to be anisotropic.

\subsection{Dissipation Rate}

In the geophysical literature, the spatial scale on which buoyant
forces become important is called the ``buoyancy scale.''  For
example,  \citet[][]{1990sidi} give it as a wavevector, expressed in terms of
the dissipation rate, $\edot$, as
\begin{equation}
  \keddy \simeq \sqrt{\frac{\ombv^3}{\edot}},
\end{equation}
Solving for the dissipation rate, this gives
\begin{equation}  
  \edot \simeq \frac{\ombv^3}{\keddy^2}.
\end{equation}
We must have $\keddy \simeq \leddy^{-1}$, so that the upper
limit on $\leddy$ in (\ref{eqn:leddy}) provides an upper limit on the
dissipation rate
\begin{equation}  \label{eqn:dissip}
  \edot \lesssim \ombv \sigv^2
\end{equation}
and a corresponding turbulent dissipation timescale

\begin{equation}  \label{eqn:theat}
  \tdiss = \frac{\sigv^2}{\edot}
  \gtrsim \frac{1}{\ombv}
  = \sqrt{\frac{\gamma}{\eta}} \frac{R}{\vkepler}.
\end{equation}
The dissipation timescale is a well-defined function of $R$ and is independent of the
turbulent speed.  Note that $\ombv^{-1}$ is comparable to the
free-fall time from $R$.  

For atmospheric turbulence driven by AGN the anisotropy is never large, so that the limits (\ref{eqn:dissip})
and (\ref{eqn:theat}) are close to identities giving the scaling for
the turbulent dissipation rate and dissipation timescales,
respectively.  Otherwise, the turbulence would need to be highly
anisotropic and predominantly transverse.  This does not seem
likely for turbulence driven by radio outbursts, but, at the minimum,
a simulation is needed to confirm it.

The main caveat on this argument is that, if $\leddy$ from equation
(\ref{eqn:leddy}) is comparable to $R$ or larger, the linear
approximation for the buoyant oscillations breaks down.  This could
occur if either the turbulent driving speed, $\sigv$, is too large compared to
the Kepler speed or $\eta$ is too small (the unperturbed atmosphere is
too nearly isentropic).  Under these conditions, vigorous stirring
would tend to mix the atmosphere radially, making it more isentropic
(assuming that the free-fall time is suitably shorter than the cooling
time).  This does not appear to be an issue for the systems observed by XRISM.  
\bibliographystyle{aasjournal}
\bibliography{biblio,bibliography}{}

\end{document}